\documentclass[11pt,a4paper]{article}

\usepackage[margin=2.5cm]{geometry}
\usepackage[T1]{fontenc}
\usepackage[utf8]{inputenc}
\usepackage{lmodern}
\usepackage{amsmath,amssymb,amsthm,mathtools}
\usepackage{graphicx}
\usepackage{booktabs,tabularx,array}
\usepackage[authoryear,round]{natbib}
\usepackage{enumitem}
\usepackage[font=small,labelfont=bf]{caption}
\usepackage{xcolor}
\usepackage[hidelinks]{hyperref}
\usepackage{setspace}

\setlist{itemsep=2pt,topsep=4pt}
\newtheorem{theorem}{Theorem}
\newtheorem{proposition}{Proposition}
\newtheorem{corollary}{Corollary}
\theoremstyle{definition}

\theoremstyle{remark}

\newcommand{\A}{\mathcal{A}}
\newcommand{\Sst}{S_{\mathrm{st}}}
\newcommand{\Smax}{S_{\max}}
\newcommand{\se}{\sigma_{e}}
\newcommand{\RF}{\mathrm{RF}}
\newcommand{\dmin}{d_{\min}}
\newcommand{\mutop}{\mu_{\mathrm{top}}}
\newcommand{\ashift}{a_{\mathrm{shift}}}
\newcommand{\aadd}{a_{\mathrm{add}}}
\newcommand{\acap}{a_{\mathrm{cap}}}
\newcommand{\diag}{\operatorname{diag}}

\title{From local competition to bump attractors:\\
how excitation width sets capacity, interference and updating\\
in recurrent models of visual working memory}

\author{Purvikalyani Prasannah\textsuperscript{1}
\and Rakesh Sengupta\textsuperscript{1,*}\\[4pt]
\small \textsuperscript{1}Computational Cognition Laboratory, School of Interwoven Arts and Sciences,\\
\small Krea University, Sri City, Andhra Pradesh, India\\
\small \textsuperscript{*}Corresponding author: \texttt{rakesh.sengupta@krea.edu.in}
(ORCID 0000-0003-1768-0551)}

\date{}

\begin{document}
\maketitle

\begin{abstract}
\noindent
On-centre off-surround (OCOS) networks and bump attractors, the two main recurrent models of visual working memory, differ chiefly in how far recurrent excitation spreads. We place them in one family parameterised by excitation width at fixed total excitation, and ask how stable each stored item is and what it costs the circuit to hold and change it. The answer is that excitation width sets the price of continuity. In the OCOS limit, which we solve exactly by completing the energy-function formalism of \citet{Sengupta2024}, an item occupies a single unit, relaxes quickly and never drifts, and capacity follows in closed form, exact over all stable states. As excitation widens towards a bump-like regime, items become continuous but costly: per item, the circuit uses 15--29 units instead of one and 6.5--11 times the activity, relaxes 35--60 times more slowly after a perturbation, needs a 35--230 times stronger cue to accept a second item, and holds 7--10 items instead of at least 40. Under noise, with background false alarms approximately matched, about a third of wide items drift, whereas narrow items never move. Yet even as an item's translation mode softens towards zero, items remain locked to the lattice. Locking is poorly predicted by energy differences between item positions, but strongly predicted by how far the unit just beyond an item's edge sits below threshold (rank correlation 0.91 across 40 parameter settings). Where that margin vanishes, a new input captures the stored item instead of being added to memory.

\end{abstract}

\section{Introduction}\label{sec:intro}

Visual working memory holds a few items for a few seconds, and it does so imperfectly. Recall becomes less precise as more items are stored \citep{BaysHusain2008,Ma2014}; beyond a handful of items, some are lost altogether \citep{LuckVogel1997,ZhangLuck2008}; nearby items interfere with one another; and a memory must stay stable against noise while remaining open to updating by new input. Whether these limits reflect a fixed number of discrete slots or a continuous resource shared among items has been debated for two decades \citep{ZhangLuck2008,BaysHusain2008,Ma2014}. At the level of neural mechanism, such properties are usually explained by recurrent networks whose persistent activity outlasts the stimulus \citep{Compte2000,Wang2001}.

It has therefore been of considerable interest to understand which recurrent mechanisms could give rise to these limits, and two families of model have dominated. In \emph{on-centre off-surround} (OCOS) networks, descended from \citet{Grossberg1973} and analysed by \citet{CohenGrossberg1983}, each unit excites itself and inhibits its neighbours. In our earlier work, an OCOS network served as a model of an attentional priority map, in which a single parameter, the strength of lateral inhibition, moved the network between a focused, capacity-limited mode and a diffuse one \citep{Sengupta2014,Sengupta2017}. The same architecture has accounted for capacity limits in both visual short-term memory and enumeration \citep{Knops2014,Sengupta2025}, and global-inhibition versions of it have been used to study interference and catastrophic loss of memories at high load \citep{EndressSzabo2017,EndressSzabo2020}. In \emph{bump-attractor} networks, by contrast, excitation spreads over a population of similarly tuned units, inhibition is broader, and a remembered feature value is held as a localised bump of activity \citep{Amari1977,Compte2000,Wimmer2014}. Bump models capture the gradual, diffusive loss of precision over a delay \citep{Compte2000,Wimmer2014}, and multi-item extensions relate their capacity limits to inhibition between bumps \citep{Edin2009,Wei2012,Almeida2015}.

The two families are usually presented as alternatives, yet they share an architecture: local excitation and broader inhibition. They differ chiefly in how far excitation reaches. This suggests treating them as members of one family, with the width $\se$ of recurrent excitation as the parameter that connects them. In the limit $\se\to 0$ excitation is confined to each unit itself, which is the OCOS network; as $\se$ grows, excitation spreads and the network approaches a bump attractor. We therefore asked a different question from the usual one: not which model is right, but what changes, and through which mechanism, as a network moves from one to the other. Two quantities frame the comparison. The first is \emph{stability}: how firmly a stored item resists perturbation, noise and new input. The second is \emph{cost}: what the circuit must spend to hold an item and to change what it holds, in units recruited, activity sustained, time to recover from a perturbation, spacing between items, and input needed to update. A memory that is cheap and stable is useful; a memory that is stable only at great cost, or cheap but fragile, is less so. Asking how stability and cost move together along the family makes the comparison between the two model types concrete.

A related literature has already shown that the discreteness of an attractor matters. A perfectly translation-invariant bump network has a continuum of stable states, and noise makes the bump wander along it \citep{Compte2000,Burak2012,KilpatrickErmentrout2013}. Breaking that symmetry, for instance with heterogeneous excitation, stabilises a discrete set of states and slows the wandering, at the cost of coarse-graining what can be stored \citep{Brody2003,Renart2003,Itskov2011,Kilpatrick2013}. These analyses, however, concern a single remembered value. Much less is known about how the same structural change shapes the multi-item properties that define visual working memory: how many items can be held, how close they can be, what happens when a new item arrives, and how memories fail under noise.

In order to address these questions, we make three contributions.
\begin{enumerate}[leftmargin=*]
\item \textbf{An exact theory of the narrow (OCOS) limit} (Section~\ref{sec:theory}). We complete the energy-function construction of \citet{Sengupta2024} for the full coupled network and show that its Hessian gives an exact test of strict linear stability (Theorem~\ref{thm:energy}). This places the node-wise and pairwise stability criteria of that work as a necessary condition and the exact one, respectively (Corollary~\ref{cor:levels}). For uniform inhibition we obtain closed-form conditions for the existence and the stability of states holding $S$ items (Theorem~\ref{thm:sym}), and we prove that every stable state has equal activations (Proposition~\ref{prop:asym}). The resulting capacity $\Sst$ is therefore exact over all stable states and always smaller than the bound $\Smax$ at which multi-item states cease to exist. We further show that loss of all items above $\Smax$ is a transient rather than an attracting state (Proposition~\ref{prop:dip}), derive the minimal separation at which two items can coexist under distance-dependent inhibition (Proposition~\ref{prop:pair}), and derive the margin that protects inactive units from recruitment (Proposition~\ref{prop:recruit}). Each result is checked against simulation.
\item \textbf{A parameterised family linking OCOS to bump attractors} (Section~\ref{sec:model}), in which excitation width varies while total excitation is held fixed, so that width is the only thing that changes.
\item \textbf{A systematic comparison of stability and cost along that family} (Section~\ref{sec:results}): local stability, resistance to being moved, the outcome of an updating cue and failure under noise, against the units, activity, relaxation time, spacing and input that each stored item requires. The comparison also yields a mechanism for lattice locking and for the capture of stored items by new input.
\end{enumerate}

In brief, excitation width sets the price of continuity. Narrow networks hold many cheap items, each occupying a single unit, locked in place, quick to recover from a perturbation and never moved by noise. Wide networks hold few items, each continuous but costly: it occupies many units, sustains several times the activity, recovers tens of times more slowly, keeps other items far away, demands a much stronger cue before a second item can be added, and drifts under noise. Along the way, the slowest mode of a stored item, which is its translation, softens towards zero, yet items stay locked to the lattice. Interestingly, locking is poorly predicted by energy differences between item positions; it is strongly predicted instead by how far the unit just beyond an item's edge sits below threshold. Where that margin vanishes, a new input no longer adds an item but captures the stored one. We close by discussing what this implies for the slots-versus-resources debate and for the circuit-level interpretation of capacity limits (Section~\ref{sec:discussion}).

\section{Model}\label{sec:model}

\subsection{The additive network}
The model uses a single layer of $N$ rate units arranged on a ring with lattice spacing 1, so that the distance between units $i$ and $j$ is $d_{ij}=\min(|i-j|,\,N-|i-j|)$. Each unit excites itself (and, along the family below, its near neighbours) and inhibits the others. The temporal dynamics of the network are given by
\begin{equation}\label{eq:model}
\dot x_i = -\lambda x_i + \sum_{j} W_{ij}\,F(x_j) + I_i(t) - \theta + \eta_i(t),
\qquad
F(x)=\begin{cases} x/(1+x), & x>0,\\ 0, & x\le 0,\end{cases}
\end{equation}
where $x_i$ is the activation of unit $i$, $\lambda>0$ is the leak, $W$ is the symmetric recurrent weight matrix, $I_i(t)$ is external input (present only while a cue is shown), $\theta\ge 0$ is an optional tonic inhibition, and $\eta_i$ is noise. The saturating, thresholded output $F$ is the one we used in earlier OCOS models of enumeration and capacity \citep{Sengupta2014,Sengupta2017,Sengupta2025}. Note that $F'(x)=(1+x)^{-2}$ for $x>0$, $F'(x)=0$ for $x<0$, and $F'(0^+)=1$. We write $y_i=F(x_i)$ and call a unit \emph{active} if $x_i>0$; the set of active units is $\A$. Unless stated otherwise, $\theta=0$ and $\eta\equiv 0$.

\subsection{A family from OCOS to bump attractors}
\begin{figure}[t]
\centering
\includegraphics[width=\textwidth]{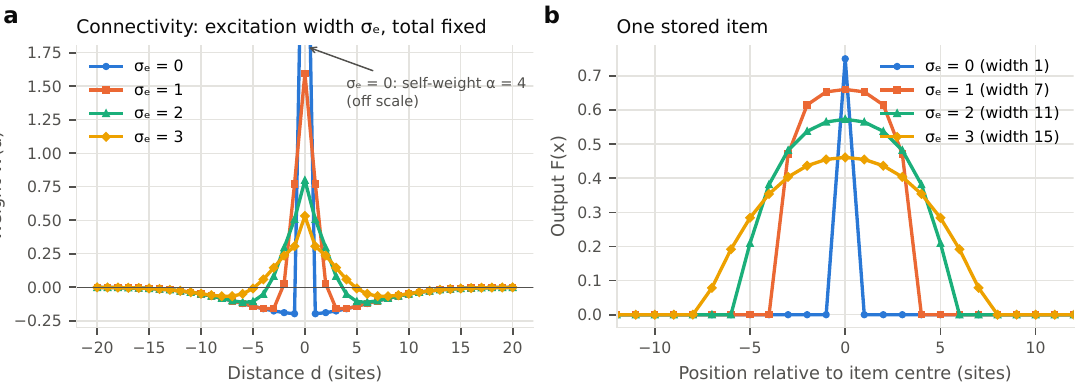}
\caption{\textbf{One family from OCOS to bump attractors.} (a)~Recurrent weight $W(d)$ as a function of distance for four excitation widths $\se$, with total excitation fixed at $\alpha=4$ ($\beta_0=0.2$, $\RF=6$). At $\se=0$ the only excitation is the self-weight $W(0)=\alpha$ (off scale). (b)~Output $F(x)$ of a single stored item for the same widths. The item occupies one unit at $\se=0$ and spreads to 7, 11 and 15 units as excitation widens.}
\label{fig:model}
\end{figure}

The recurrent weights are excitation minus inhibition, $W=E-B$, with
\begin{equation}\label{eq:kernels}
E_{ij}=\alpha\,\frac{g_e(d_{ij})}{\sum_{d} g_e(d)},\quad g_e(d)=e^{-d^2/2\se^2},
\qquad
B_{ij}=\beta_0\, e^{-d_{ij}^2/2\RF^2}\ (i\ne j),\quad B_{ii}=0 .
\end{equation}
The inhibitory kernel $B$, with strength $\beta_0$ and width $\RF$, is the distance-dependent inhibition of \citet{Sengupta2025}. The excitatory kernel $E$ has width $\se$ and is normalised so that the \emph{total} excitation each unit sends, $\sum_j E_{ij}=\alpha$, does not depend on $\se$. This ``mass'' normalisation is the key design choice: it isolates width as the only thing that changes along the family. At $\se=0$ the kernel reduces to $E_{ij}=\alpha\,\delta_{ij}$ and \eqref{eq:model} is exactly the OCOS network
\begin{equation}\label{eq:ocos}
\dot x_i=-\lambda x_i+\alpha F(x_i)-\sum_{j\ne i}\beta_{ij}F(x_j)+I_i,
\end{equation}
with $\beta_{ij}=B_{ij}$. As $\se$ grows, excitation spreads to neighbours, an item comes to occupy several units, and the network acquires the local-excitation, broad-inhibition structure of a bump attractor \citep{Amari1977,Compte2000}. Figure~\ref{fig:model} shows the kernel and a stored item at several widths. In a control (Section~\ref{sec:controls}) we use instead a ``peak'' normalisation, $E_{ij}=\alpha\,g_e(d_{ij})$, in which the self-weight stays at $\alpha$ and total excitation grows with $\se$.

The OCOS limit with uniform inhibition, $\beta_{ij}=\beta$ for all $i\ne j$, is the global-inhibition network studied by \citet{EndressSzabo2020} and, with a single inhibition parameter, the network of our earlier enumeration work \citep{Sengupta2014,Sengupta2017}; it is the setting for the closed-form results of Section~\ref{sec:theory}.

\paragraph{Model behaviour.} Self-excitation and lateral inhibition create a push--pull interaction that drives the network to a stable equilibrium. A cued unit is held above threshold by its own recurrent excitation, while the inhibition it sends limits how many other units can remain active alongside it. At $\se=0$ an item is therefore a single active unit, and the number of items the network can hold is set by inhibition alone. As excitation widens, neighbouring units begin to support one another, so that an item becomes a small group of co-active units whose position can, in principle, shift without the item being lost. The question pursued in the rest of the paper is how far along the family this second, bump-like behaviour actually emerges, and what it costs.

\subsection{Simulations and measures}\label{sec:measures}
Equation~\eqref{eq:model} was integrated with the Euler (noise-free) or Euler--Maruyama method with step $0.01$. An item was created by a cue of amplitude 1 at one unit for 5 time units, followed by 60 time units without input. In the course of the simulations, two numerical safeguards proved essential. First, a cue at a single unit is mirror-symmetric, and a symmetric state can be an unstable saddle on which the dynamics never leaves the symmetric subspace; for example, an item centred on a unit whose stable form is centred between two units. Second, near the bump-attractor end the slowest relaxation time $1/|\mutop|$ can exceed the settling time. Every final state was therefore tested for convergence (largest residual $<10^{-3}$) and linear stability; states failing the test were perturbed slightly along their leading eigenvector, restricted to active units, and allowed to settle again (Appendix~\ref{app:methods}). All quantities below refer to states that passed.

For each pair $(\se,\beta_0)$ we measured the following quantities.
\begin{description}[leftmargin=1.2em,font=\normalfont\itshape]
\item[Width:] the number of active units in a single stored item.
\item[Slowest mode $\mutop$:] the largest real part of the Jacobian's eigenvalues at that item. Negative values close to zero mean a nearly flat direction in the energy landscape.
\item[Locking $\ashift$:] the smallest input, applied for 5 time units to the first unit beyond the item's edge, that leaves a single item displaced by at least half a site. It measures how strongly the lattice holds an item in place.
\item[Interference range $\dmin$:] the smallest separation from which two items, cued together with slightly unequal amplitudes (1 and 1.02), never merge or lose an item.
\item[Capacity:] the largest $S$ for which $S$ equally spaced items, cued together, all survive in a stable state (searched up to $S=40$).
\item[Updating:] a cue at a fixed distance $2\RF$ from a stored item, applied for 10 time units, over 25 log-spaced amplitudes from 0.005 to 5. Each outcome is classified as \emph{ignored} (one item, unmoved), \emph{pushed} (one item, displaced but not onto the cue), \emph{added} (two stable items), or \emph{captured} (one item, now at the cue). $\aadd$ and $\acap$ are the smallest amplitudes giving each outcome.
\item[Noise:] trials of 100 time units with independent white noise of intensity $\sigma$ added to every unit, classified as \emph{held}, \emph{moved} (displaced by at least 0.75 sites), \emph{intrusion} (a spurious second item) or \emph{lost}. Noise runs include tonic inhibition $\theta$, without which distant quiescent units have no protection against recruitment (Section~\ref{sec:recruit}). The main noise comparison (Section~\ref{sec:noise}) uses 1000 trials per condition and compares widths both at a fixed $\theta$ and at a $\theta$ calibrated to equalise background false alarms.
\end{description}

Two families of runs explore the continuum, both with total excitation fixed at $\alpha=4$ and $\lambda=1$: run~A ($N=160$, $\RF=6$, $\beta_0\in\{0.2,0.3\}$, $\se\in\{0,0.5,\dots,3\}$) and run~B ($N=320$, $\RF=12$, $\beta_0\in\{0.1,0.15\}$, $\se\in\{0,1,2,3,4,6\}$). Run~B doubles both the ring and the inhibitory width, and halves $\beta_0$ so that total inhibition matches run~A; it tests whether the trends continue at larger relative widths. Two control runs (Section~\ref{sec:controls}) use $\alpha=2.2$, the value of \citet{EndressSzabo2020}: run~C with mass normalisation and run~D with peak normalisation. All parameters are listed in Table~\ref{tab:params} (Appendix~\ref{app:methods}).

\section{Theory of the narrow limit}\label{sec:theory}

Before exploring the family numerically, we derive what can be said exactly. The first result below holds for any symmetric $W$ and therefore along the whole family; the rest concern the OCOS limit \eqref{eq:ocos}, where the network is simple enough for closed forms. Throughout, ``stable'' means strictly linearly stable in the sense of Theorem~\ref{thm:energy}(b); at the boundaries of the conditions below, where $H$ is singular, the linearisation is inconclusive. Proofs are in Appendix~\ref{app:proofs}.

\subsection{The energy function and exact stability}

Energy functions have played two roles in our earlier work on OCOS networks: as a source of reaction-time correlates for enumeration \citep{Sengupta2014,Sengupta2017}, and as a tool for deriving stability conditions \citep{Sengupta2024}. In the latter, energy functions for Cohen--Grossberg networks were constructed from the dynamics themselves, rather than by guessing a Lyapunov function: for each unit, $dH_i\propto\dot x_i\,d\dot x_i$. Near equilibrium this construction recovers the Cohen--Grossberg function \citep{CohenGrossberg1983}. For the additive OCOS network, keeping each unit's own dependence of $\dot x_i$ on $x_i$, it gives the node-wise energy
\begin{equation}\label{eq:sbp}
H\;\propto\;-\sum_i\int\big(\lambda-\alpha F'(x_i)\big)\,\dot x_i^{\,2}\,dt ,
\end{equation}
written here for general $\lambda$ (they used $\lambda=1$). At a state with $S$ equally active units, the coefficient $\lambda-\alpha F'(x_i)$ is positive when $A_S^2>\lambda\alpha$, with $A_S=\alpha-(S-1)\beta$. \citet{Sengupta2024} also derived a second condition, $A_S^2>\lambda(\alpha+\beta)$, from the decay of differences between pairs of units, and noted that the two differ, increasingly so as $\beta$ grows. This left open which of the two is the right one. We now complete the construction for the full coupled network, which settles exactly how the two conditions relate.

\begin{theorem}[Energy and Hessian test]\label{thm:energy}
Let $W$ be symmetric, $I$ constant and $\eta\equiv 0$ (a constant $\theta$ is absorbed into $I$). Let $\Phi(x)=\ln(1+x)+\frac{1}{1+x}-1$ for $x>0$ and $\Phi(x)=0$ for $x\le 0$, and
\begin{equation}\label{eq:energy}
\mathcal{E}(x)=\sum_i\Big[\lambda\,\Phi(x_i)-I_iF(x_i)\Big]-\tfrac12\sum_{i,j}W_{ij}F(x_i)F(x_j).
\end{equation}
\begin{enumerate}[label=(\alph*)]
\item Along solutions of \eqref{eq:model}, $\dot{\mathcal E}=-\sum_i F'(x_i)\,\dot x_i^{\,2}\le 0$.
\item Let $x^*$ be an equilibrium with no unit exactly at threshold, and let
\begin{equation}\label{eq:hessian}
H_{ij}=\lambda(1+x^*_i)^2\,\delta_{ij}-W_{ij},\qquad i,j\in\A,
\end{equation}
be the Hessian of $\mathcal E$ in the output coordinates $y_i=F(x_i)$. All eigenvalues of the Jacobian at $x^*$ are real. If $H\succ 0$, $x^*$ is linearly, and hence locally exponentially, stable. If $H$ has a negative eigenvalue, $x^*$ is linearly unstable. If $H$ is singular, linearisation is inconclusive.
\end{enumerate}
\end{theorem}

Part~(a) is the dissipation law that the node-wise construction encodes. Each unit's contribution to $\dot{\mathcal E}$ is weighted by $F'(x_i)$, as in \eqref{eq:sbp}. Part~(b) makes the energy a precise tool: strict linear stability of a memory state is equivalent to its being a nondegenerate local minimum of $\mathcal E$. For the OCOS network, $H_{ii}=\lambda(1+x_i^*)^2-\alpha$ and $H_{ij}=\beta_{ij}$. Since inactive units have $F'=0$, they never feed back on active ones, and only units that are active together enter $H$.

\begin{corollary}[Criteria for strict linear stability]\label{cor:levels}
At an equilibrium of the OCOS network with no unit at threshold:
\begin{enumerate}[label=(\roman*)]
\item \emph{Necessary:} every node-wise coefficient of \eqref{eq:sbp} is positive, $\lambda-\alpha F'(x_i^*)>0$; equivalently, every diagonal entry of $H$ is positive.
\item \emph{Sufficient:} $\lambda(1+x_i^*)^2-\alpha>\sum_{j\in\A,\,j\ne i}\beta_{ij}$ for every active $i$ (Gershgorin's theorem applied to $H$).
\item \emph{Exact (strict linear stability):} $H\succ 0$.
\end{enumerate}
For $S\ge 2$ equally active units under uniform inhibition, (i) reads $A_S^2>\lambda\alpha$ and (iii) reads $A_S^2>\lambda(\alpha+\beta)$. All three concern strict linear stability; where $H$ is singular, or a unit sits at threshold, they do not decide nonlinear stability.
\end{corollary}

Hence, the node-wise energy criterion of \citet{Sengupta2024} is necessary but not sufficient, and the pairwise criterion is the exact one (Theorem~\ref{thm:sym} below). The difference between them is the inhibitory coupling $\beta$ between co-active units, which the node-wise construction leaves out; this accounts for the gap observed there. In (ii), only co-active units enter the sum: inactive neighbours are irrelevant, however many lie within the inhibitory range.

The energy framework also provides a useful perspective on two properties that matter below. First, the energy is \emph{blind below threshold}: inactive units contribute nothing to $\mathcal E$ (since $\Phi=F=0$ there), and their weight in the dissipation law is $F'=0$. The energy therefore governs how active units settle within a given set of active units, but not what it takes to change that set. That depends on how far an inactive unit sits below threshold (Section~\ref{sec:recruit}), and it is the quantity that best predicts how strongly items are locked to the lattice (Section~\ref{sec:locking}). Second, the active Jacobian is $-HD$ with $D=\diag(F'(x_i))$, which has the same eigenvalues as the symmetric matrix $-D^{1/2}HD^{1/2}$. So $\mutop\to 0^-$ means that $H$ is becoming singular: the energy landscape is flattening along some direction. For a translation-invariant bump attractor that direction is translation, and $\mutop=0$ exactly \citep{Amari1977,Compte2000}. Section~\ref{sec:continuum} checks which direction is flattening here.

\begin{proposition}[Persistent memory requires $\alpha>\lambda$]\label{prop:alpha}
For the OCOS network with $I=0$, any equilibrium with an active unit satisfies $\alpha>\lambda$. Conversely, if $\alpha>\lambda$, a single active unit at $x_w=\alpha/\lambda-1$, with all other units inactive, is a stable equilibrium.
\end{proposition}

\subsection{Multi-item states and capacity}

\begin{theorem}[Existence, stability and capacity]\label{thm:sym}
Consider the OCOS network with uniform inhibition $\beta_{ij}=\beta$ and $I=0$. Let $A_S=\alpha-(S-1)\beta$. A state with $S$ equally active units and the rest inactive is
\begin{equation}
x_S=\frac{A_S}{\lambda}-1,\qquad x_l=-\frac{\beta S y_S}{\lambda}\ \ (l\notin\A),\qquad y_S=1-\frac{\lambda}{A_S}.
\end{equation}
\begin{enumerate}[label=(\alph*)]
\item It exists if and only if $A_S>\lambda$, that is, $S<\Smax=1+(\alpha-\lambda)/\beta$.
\item Its active Jacobian has a uniform mode and, for $S\ge 2$, an $(S-1)$-fold transverse mode,
\begin{equation}\label{eq:modes}
\mu_\parallel=-\lambda+\frac{\lambda^2}{A_S},\qquad
\mu_\perp=-\lambda+\frac{\lambda^2(\alpha+\beta)}{A_S^2}.
\end{equation}
\item For $S\ge 2$ it is stable if and only if $A_S^2>\lambda(\alpha+\beta)$, that is,
\begin{equation}\label{eq:Sst}
S<\Sst=1+\frac{\alpha-\sqrt{\lambda(\alpha+\beta)}}{\beta},
\end{equation}
and $\Sst<\Smax$ always. Equivalently, at fixed $\beta$ the $S$-item state ($S\ge 2$) is stable if and only if
\begin{equation}\label{eq:alphaS}
\alpha>\alpha_S(\beta)=\tfrac12\Big[2(S-1)\beta+\lambda+\sqrt{\lambda^2+4\lambda S\beta}\Big].
\end{equation}
\end{enumerate}
A single item ($S=1$) has no transverse mode and is stable whenever it exists, that is, whenever $\alpha>\lambda$.
\end{theorem}

The capacity of the network, the largest number of items it can hold stably, is therefore $\max(1,\lceil\Sst\rceil-1)$ whenever $\alpha>\lambda$: the largest integer strictly below $\Sst$, but never less than one. This is the largest number of simultaneously active units in a strictly linearly stable equilibrium. How many items a particular cueing protocol actually recovers can be smaller, a distinction that matters for the spatially local networks of Section~\ref{sec:continuum}. Between $\Sst$ and $\Smax$ the $S$-item state exists but is a saddle: it can be reached from exactly symmetric cues, but any asymmetry makes the network shed items until a stable state is reached. In other words, the push--pull balance between self-excitation and shared inhibition can be struck for more items than it can be maintained against small perturbations. With $\alpha=2.2$ and $\lambda=1$ \citep{EndressSzabo2020}, $\beta=0.1$ gives $\Sst=7.83$ (capacity 7) against $\Smax=13$, and $\beta=0.15$ gives $\Sst=5.45$ (capacity 5) against $\Smax=9$.

\begin{corollary}[Winner-take-all window]\label{cor:wta}
Only single-item states are stable if and only if $\lambda<\alpha\le\alpha_2(\beta)=\tfrac12\big[\lambda+2\beta+\sqrt{\lambda^2+8\lambda\beta}\big]$. Above $\alpha_2$ the network stores two or more items; the single-item state remains stable for every $\alpha>\lambda$.
\end{corollary}

\noindent The lower bound $\alpha>\lambda$ coincides with the one obtained from the node-wise energy at $n=1$ in \citet{Sengupta2024}. For the upper bound we reported there $\alpha^2-\alpha-2\beta\le 0$ (with $\lambda=1$). Expanding the pairwise criterion $(\alpha-\beta)^2\le\alpha+\beta$ exactly gives Corollary~\ref{cor:wta} instead; for example, the upper bound at $\beta=0.25$ is $1.62$ rather than $1.37$. The winner-take-all window is therefore wider than previously stated.

\begin{proposition}[Stable states have equal activations]\label{prop:asym}
Under uniform inhibition with $I=0$, every strictly linearly stable equilibrium with no unit at threshold has all its active units at the same activation.
\end{proposition}

\noindent Memory states under uniform inhibition are therefore indexed by the number of items alone. Together with Theorem~\ref{thm:sym}, Proposition~\ref{prop:asym} makes $\max(1,\lceil\Sst\rceil-1)$ the capacity over \emph{all} strictly linearly stable equilibria, not only those with equal activations.

\subsection{Above $\Smax$: total loss is transient}

When more than $\Smax$ items are cued simultaneously, the $S$-item state does not exist, and an exactly symmetric configuration decays towards silence. This is the catastrophic interference of global-inhibition networks \citep{EndressSzabo2020}. With this activation function, however, silence is an unstable equilibrium: a trajectory that starts exactly at silence stays there, but silence does not attract nearby states, since any small positive perturbation grows when $\alpha>\lambda$.

\begin{proposition}[Transient dip above $\Smax$, asymptotic for small $\varepsilon$]\label{prop:dip}
Let $\beta_{ij}=\beta$, $I=0$, $\alpha>\lambda$ and $S>\Smax$, and start from $S$ nearly equal active units with small relative asymmetry $\varepsilon$ that enter the regime $x\ll 1$ before the asymmetry has grown. Near $x=0^+$ the mean activity decays at rate $\lambda-A_S>0$ while differences grow at rate $\alpha+\beta-\lambda>0$, so the asymmetry relative to the mean grows at rate $S\beta$. Activity therefore reaches a minimum, to leading order as $\varepsilon\to 0$, at
\begin{equation}\label{eq:dip}
t_{\mathrm{dip}}\simeq t_0+\frac{\ln(1/\varepsilon)}{S\beta},\qquad
\text{depth}\propto\varepsilon^{(\lambda-A_S)/(S\beta)},
\end{equation}
after which a small number of items recovers. Only an exactly symmetric start decays to silence.
\end{proposition}

At $S=\Smax$ exactly, the uniform mode is marginal ($\lambda-A_S=0$): the mean then decays algebraically, $x\approx 1/(\lambda t)$ for small $x$, and \eqref{eq:dip} does not apply. Within a finite trial, apparent total loss therefore depends on how unequal the cues are as well as on how many there are. Loss is observed when $t_{\mathrm{dip}}$ exceeds the delay or the dip falls below the read-out threshold. The result rests on $F'(0^+)=1$ together with $\alpha>\lambda$, which make silence unstable to positive perturbations; with a firing threshold inside $F$, silence could be attracting. Shunting (multiplicative) gating does not remove the underlying existence limit: the shunting OCOS network has its own bound on the number of equally active units, and its energy criterion again coincides with linear stability (Appendix~\ref{app:shunting}).

\subsection{Two items under distance-dependent inhibition}

\begin{proposition}[Minimal separation]\label{prop:pair}
In the OCOS network with $I=0$, two items at units $i$ and $j$, with no other units active, form the equilibrium $x_i=x_j=(\alpha-\beta_{ij})/\lambda-1$. It is stable if and only if
\begin{equation}\label{eq:betastar}
\beta_{ij}<\beta^*=\tfrac12\Big[(2\alpha+\lambda)-\sqrt{\lambda^2+8\alpha\lambda}\Big].
\end{equation}
With the Gaussian kernel \eqref{eq:kernels}, two items therefore coexist if and only if they are farther apart than
\begin{equation}\label{eq:dstar}
d^*=\RF\sqrt{2\ln(\beta_0/\beta^*)}\qquad(\text{at any separation if }\beta_0\le\beta^*).
\end{equation}
On the lattice, the smallest admissible separation is therefore $\dmin=\max\{1,\lfloor d^*\rfloor+1\}$ when $\beta_0>\beta^*$, and $\dmin=1$ otherwise.
\end{proposition}

\noindent This turns the qualitative idea that interference depends on local rather than global crowding into a closed-form law for the OCOS limit: two items interfere only if the inhibition between them exceeds what self-excitation can withstand. Section~\ref{sec:results} measures how the separation two items need changes as excitation widens.

\subsection{Protection of quiescent units}\label{sec:recruit}

\begin{proposition}[Recruitment barrier]\label{prop:recruit}
Consider an inactive unit $l$ of the OCOS network and treat the inhibition it receives as frozen while it responds (a fixed-background approximation): $m_l=\sum_j\beta_{lj}y_j-I_l>0$, so that it rests at $x_l=-m_l/\lambda$. Let $h(x)=-\lambda x+\alpha F(x)$, whose maximum on $x>0$ is $h_{\max}=(\sqrt{\alpha}-\sqrt{\lambda})^2$.
\begin{enumerate}[label=(\alph*)]
\item If $m_l\ge h_{\max}$, no transient kick can recruit unit $l$ while its inhibition stays fixed.
\item If $m_l<h_{\max}$, any kick larger than $\Delta_l=m_l/\lambda+x_u$ recruits it, where $x_u$ is the smaller positive root of $h(x)=m_l$.
\item A unit receiving no inhibition ($m_l=0$) has $\Delta_l=0$ when $\alpha>\lambda$, so it is unprotected.
\end{enumerate}
\end{proposition}

\noindent Because the background is frozen, $\Delta_l$ is a local threshold for the unit itself, not an exact barrier for the whole network under arbitrary perturbations: once unit $l$ responds, the inhibition it receives can change. Figure~\ref{fig:theory}d shows that it is nonetheless accurate for single kicks. The same margin generalises to $\se>0$ as $m_l=-\sum_jW_{lj}y_j-I_l$, the net drive that holds an inactive unit below threshold. For the unit just beyond an item's edge, it measures what it takes to change the item's active set; Section~\ref{sec:locking} shows that it predicts an item's resistance to displacement.

Part~(c) matters for every network with local inhibition: units far from all stored items receive almost no inhibition, so any positive fluctuation can grow into a spurious item. Some background-quieting mechanism is required, such as a firing threshold, tonic inhibition or a weak global inhibitory component. Our noise simulations use tonic inhibition $\theta$, which shifts every quiescent unit's margin by $\theta$.

\subsection{Numerical confirmation}

\begin{figure}[t]
\centering
\includegraphics[width=\textwidth]{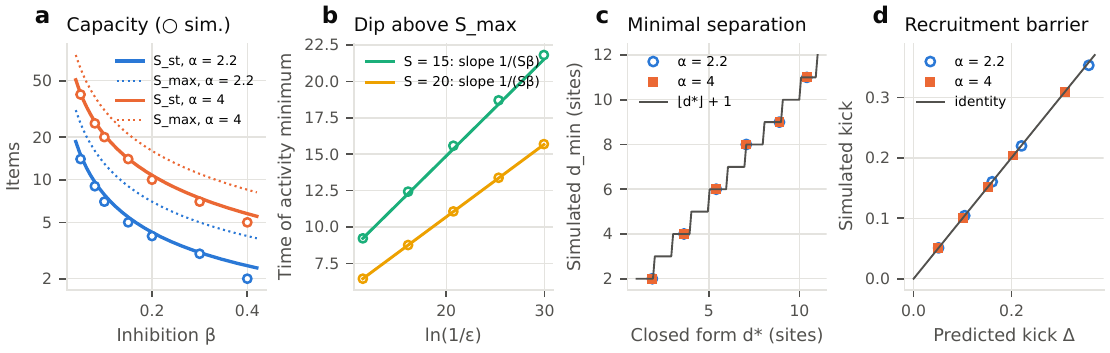}
\caption{\textbf{Closed forms for the OCOS limit agree with simulation.} (a)~Capacity under uniform inhibition: $\Sst$ (solid) and $\Smax$ (dotted) for $\alpha=2.2$ and $\alpha=4$, with simulated capacity (circles). (b)~Above $\Smax$ ($\alpha=2.2$, $\beta=0.1$): time of the activity minimum against $\ln(1/\varepsilon)$ for $S=15$ and $S=20$; lines have the predicted slope $1/(S\beta)$. (c)~Simulated minimal separation of two items against the closed form $d^*$ ($\RF=6$; line: $\lfloor d^*\rfloor+1$). (d)~Simulated recruitment kick against the predicted barrier $\Delta$.}
\label{fig:theory}
\end{figure}

We validated each closed form through simulation (Fig.~\ref{fig:theory}). The simulated capacity under uniform inhibition equalled $\lceil\Sst\rceil-1$ in all 14 cases tested (two values of $\alpha$, seven of $\beta$). After a dip above $\Smax$, the time of the activity minimum grew with $\ln(1/\varepsilon)$ at slopes of 0.502 and 0.683 for $S=20$ and $S=15$, against predicted $1/(S\beta)$ values of 0.500 and 0.667; the depth exponents were 0.353 and 0.147 against 0.350 and 0.133. The approximation is closer for larger $S$, where the dip reaches lower activity. The simulated minimal separation equalled $\lfloor d^*\rfloor+1$ in all 12 cases, and the simulated recruitment kick agreed with $\Delta$ to within 1.4\%.

\section{Results: from OCOS to bump attractors}\label{sec:results}

\subsection{The simulations reproduce the narrow limit exactly}

Before varying excitation width, we first asked whether the simulation code reproduces the closed forms of Section~\ref{sec:theory} at $\se=0$, where the answer is known exactly (Table~\ref{tab:checks}). The vector field coincides with an independent implementation of \eqref{eq:ocos}; capacity under uniform inhibition matches Theorem~\ref{thm:sym}; the minimal separation matches Proposition~\ref{prop:pair} both when it is nontrivial and when any separation suffices; the cue needed to add a second item matches the inhibitory margin at the cued site; and the slowest mode of a single item equals the analytic value $-\lambda+\lambda^2/\alpha$. The capacity check also shows the shedding predicted by Theorem~\ref{thm:sym}: eight and nine simultaneous cues ended with seven and six items.

\begin{table}[t]
\centering\small
\caption{Checks at $\se=0$, run automatically before every sweep ($\alpha=2.2$, $\lambda=1$, $N=160$, $\RF=6$).}
\label{tab:checks}
\begin{tabularx}{\textwidth}{@{}l X l l@{}}
\toprule
 & Quantity & Simulated & Closed form \\
\midrule
C1 & Vector field vs.\ independent OCOS implementation & max.\ difference $<10^{-12}$ & identical \\
C2 & Capacity, uniform inhibition $\beta=0.1$ & 7 & $\lceil\Sst\rceil-1=7$ ($\Sst=7.83$) \\
C3 & Minimal separation, $\beta_0=1.0$ / $\beta_0=0.3$ & 7 / 1 & $\lfloor d^*\rfloor+1=7$ ($d^*=6.62$) / 1 \\
C4 & Cue to add a 2nd item at $2\RF$ / inhibitory margin & ratio 1.003 & 1 (from above) \\
C5 & Slowest mode of one item & $-0.545455$ & $-\lambda+\lambda^2/\alpha=-0.545455$ \\
\bottomrule
\end{tabularx}
\end{table}

\subsection{Widening excitation approaches a weakly stabilised, bump-like regime}\label{sec:continuum}

\begin{figure}[t]
\centering
\includegraphics[width=\textwidth]{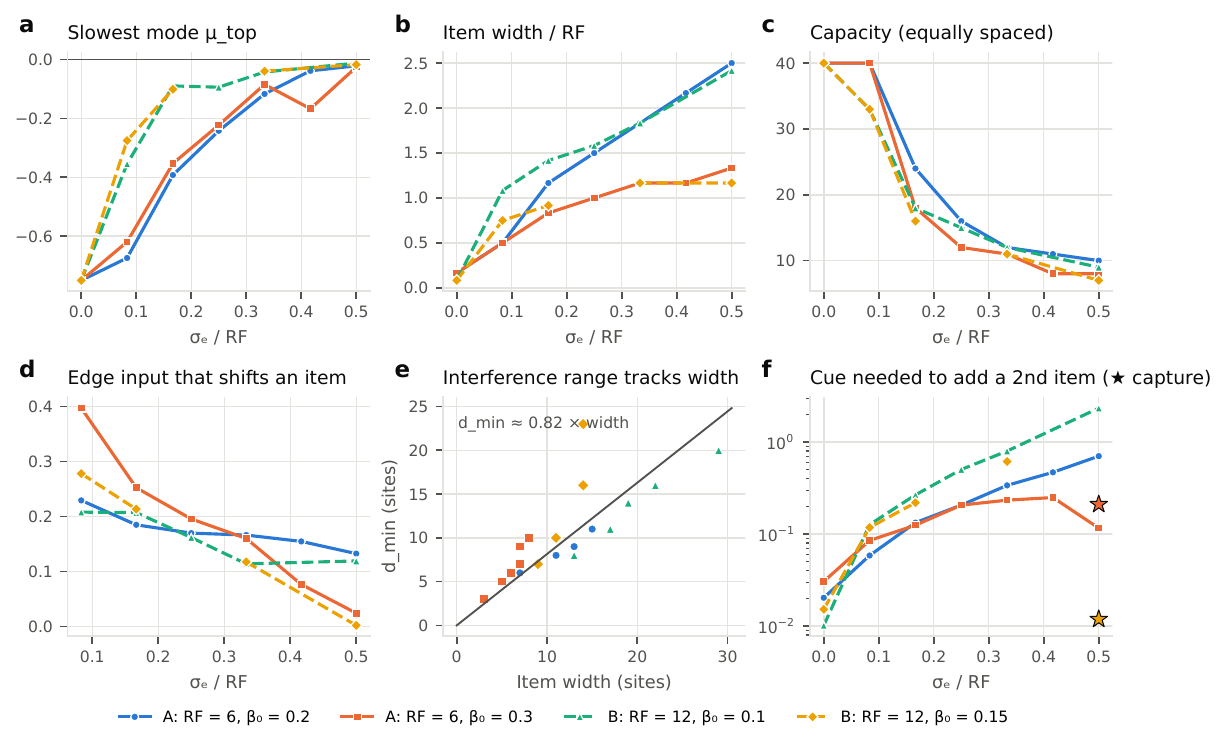}
\caption{\textbf{Widening excitation at fixed total excitation moves the OCOS network towards a bump attractor} ($\alpha=4$). Runs A ($\RF=6$, solid) and B ($\RF=12$, dashed), with width expressed relative to the inhibitory range. (a)~Slowest mode $\mutop$ of a single item. (b)~Item width relative to $\RF$. (c)~Capacity for equally spaced items (40 is the search limit). (d)~Locking: the edge input needed to shift an item; at $\se=0$ with $\beta_0<\beta^*$ an adjacent input adds an item instead, so no shift exists. (e)~Minimal separation against item width across all cells (line: least-squares fit through the origin). (f)~Cue needed to add a second item at distance $2\RF$; stars mark the cue that captures the stored item. Gaps are cells without a stable single item.}
\label{fig:continuum}
\end{figure}

Having established the narrow limit, we widened excitation at fixed total excitation and followed a stored item and pairs of items along the way (Fig.~\ref{fig:continuum}; Table~\ref{tab:continuum}). To compare runs A and B, widths are expressed relative to the inhibitory range, $\se/\RF$.

\paragraph{The slowest mode approaches zero, and it is translation.} At $\se=0$ a single item has $\mutop=-0.75$ in every run, the OCOS value $-\lambda+\lambda^2/\alpha$ for $\alpha=4$. As excitation widens, $\mutop$ rises towards zero, reaching $-0.022$ and $-0.026$ in run~A at $\se/\RF=0.5$ and $-0.012$ and $-0.017$ in run~B at the same relative width (Fig.~\ref{fig:continuum}a). By Theorem~\ref{thm:energy}, the energy landscape around an item is flattening along one direction. A vanishing eigenvalue need not mean translation, however; it could equally signal an impending bifurcation of the item's amplitude or width. We therefore checked which direction is flattening. We measured the overlap $Q$ between the slowest eigenvector and the translation vector of the item's profile (its discrete derivative), restricted to active units; $Q=1$ means the slow mode is a pure shift. In run~A, $Q\ge 0.979$ in every cell with $\se>0$, and in run~B $Q\ge 0.972$ in every such cell but one, the cell in which the edge unit sits at threshold (Section~\ref{sec:locking}). Thus, the softening mode is the item's translation, as in a bump attractor. Translation was the slowest mode even in strongly locked items, including those of run~D with $\beta_0=1$ ($Q\ge 0.988$). In fact, what changes with width is not which mode is slowest but how soft it is. This is a finite lattice of 160 or 320 units, and we do not claim convergence to the continuum neural-field limit; the claim is the narrower one that the translation mode softens towards zero. The approach is not perfectly monotone. Run~A with $\beta_0=0.3$ shows a step back at $\se=2.5$ ($\mutop=-0.167$), where item width stalls at seven units. Run~B with $\beta_0=0.15$ has no stable single item at $\se=3$ ($\mutop=+0.010$). These irregularities coincide with changes in the number of units an item occupies, which on a lattice must change in whole units.

\paragraph{Items widen, and the interference range follows.} As expected, item width grows from one unit to 15 (run~A, $\beta_0=0.2$) and 29 (run~B, $\beta_0=0.1$) units, and more slowly under stronger inhibition (Fig.~\ref{fig:continuum}b). The separation two items need grows with it: from 1 to 11 and 10 sites in run~A and from 1 to 20 and 23 sites in run~B. Across all 21 cells with items wider than one unit, $\dmin$ tended to grow in proportion to item width (correlation 0.80; least-squares slope 0.82; Fig.~\ref{fig:continuum}e), although the slope is specific to these parameters and should not be read as a general scaling law. Hence, the interference range of the OCOS limit, set by inhibition alone (Proposition~\ref{prop:pair}), gives way to a range set mainly by how far each item extends.

\paragraph{Operational capacity falls.} We measured capacity operationally, as the largest number of equally spaced items, cued together, that all survive in a stable state; this depends on the loading protocol and the ring size and is not the largest number of items any stable state could hold. It was at least 40 items (the search limit) at $\se=0$ in every run, and fell to 10 and 8 in run~A and 9 and 7 in run~B at $\se/\RF=0.5$ (Fig.~\ref{fig:continuum}c). On a ring of fixed size, capacity under local inhibition is limited by how many items fit at their mutual interference range, so the fall in capacity largely mirrors the rise in $\dmin$. These are items on an artificial one-dimensional lattice and are not comparable to behavioural estimates of three or four items.

\begin{table}[t]
\centering\small
\caption{Stability and cost of a stored item along the continuum (runs A, $\beta_0=0.2$, and B, $\beta_0=0.1$; $\alpha=4$). Width: active units per item; activity: total output $\sum_iF(x_i)$ of the item; $\tau=1/|\mutop|$: slowest relaxation time. Capacity at $\se=0$ reached the search limit of 40.}
\label{tab:continuum}
\begin{tabular}{@{}l r r r r r r r r r@{}}
\toprule
 & & \multicolumn{2}{c}{Stability} & \multicolumn{6}{c}{Cost} \\
\cmidrule(lr){3-4}\cmidrule(l){5-10}
 & $\se$ & $\mutop$ & $\ashift$ & width & activity & $\tau$ & $\dmin$ & capacity & $\aadd$ \\
\midrule
Run A & 0 & $-0.750$ & -- & 1 & 0.75 & 1.3 & 1 & $\ge 40$ & 0.020 \\
($\RF=6$) & 1 & $-0.392$ & 0.185 & 7 & 4.14 & 2.5 & 6 & 24 & 0.133 \\
 & 2 & $-0.117$ & 0.166 & 11 & 4.92 & 8.5 & 8 & 12 & 0.339 \\
 & 3 & $-0.022$ & 0.132 & 15 & 4.87 & 46.3 & 11 & 10 & 0.703 \\
\midrule
Run B & 0 & $-0.750$ & -- & 1 & 0.75 & 1.3 & 1 & $\ge 40$ & 0.010 \\
($\RF=12$) & 2 & $-0.090$ & 0.207 & 17 & 8.96 & 11.1 & 11 & 18 & 0.267 \\
 & 4 & $-0.042$ & 0.114 & 22 & 9.07 & 23.6 & 16 & 12 & 0.798 \\
 & 6 & $-0.012$ & 0.119 & 29 & 8.48 & 80.5 & 20 & 9 & 2.338 \\
\bottomrule
\end{tabular}
\end{table}

\subsection{What continuity costs}\label{sec:cost}

The bump-like regime is reached at a price. To make that price explicit, we measured what the circuit must spend on a single stored item (Table~\ref{tab:continuum}; Appendix~\ref{app:methods}). We consider four costs. The first is neural resources: an item occupies one unit at $\se=0$ and 15 (run~A) or 29 (run~B) units at $\se/\RF=0.5$. The second is activity: the item's total output $\sum_iF(x_i)$, a rate-based proxy for the activity the circuit must sustain, rises from 0.75 to 4.87 and 8.48, that is 6.5 and 11 times the OCOS value, and does so mostly over the first steps of widening. The third is time: the slowest relaxation time $\tau=1/|\mutop|$, which sets how long an item takes to recover from a perturbation, grows from 1.3 to 46 and 80 time units (35 and 60 times). The fourth is the cost of sharing and changing the store: the separation that other items must keep grows from 1 to 11 and 20 sites, so that operational capacity falls, and the cue needed to add a second item grows 35 and 230 times (Section~\ref{sec:updating}). On each of these axes the narrow network is the cheaper one, and it is also the more stable one: it recovers faster from a perturbation (Section~\ref{sec:continuum}) and is never moved by noise (Section~\ref{sec:noise}). What widening buys is continuity: an item whose position can shift smoothly rather than by whole units.

These are costs to the circuit, not to the simulation. Because the recurrent sum in \eqref{eq:model} is a convolution on the ring, one integration step costs $O(N\log N)$ operations with fast Fourier transforms, whatever the width, so the cost of simulating the network does not distinguish the two regimes. The cost of \emph{certifying} a memory state does differ. In the OCOS limit with uniform inhibition the stability of an $S$-item state is decided by a closed-form inequality (Theorem~\ref{thm:sym}). Along the rest of the family, Theorem~\ref{thm:energy} requires the spectrum of the Hessian over the whole active set, whose size grows with the number of units per item. The activity measure, likewise, is a proxy and not metabolic energy, which would require assumptions about spike generation and synaptic events that a rate model does not make.

\subsection{Lattice locking is predicted by the recruitment margin, not by energy differences}\label{sec:locking}

A continuous attractor offers no resistance to being moved. If widening simply turned items into bumps, resistance to displacement should therefore vanish together with $\mutop$. Interestingly, it did not. Across runs A and B, the edge input needed to shift an item stayed between 0.11 and 0.40 in all but three cells with $\se>0$ (Fig.~\ref{fig:continuum}d), even where $\mutop$ was within a few hundredths of zero. One cell was weakly locked (run~A, $\beta_0=0.3$, $\se=2.5$; $\ashift=0.076$), and two lost their locking almost entirely: run~A with $\beta_0=0.3$ at $\se=3$ ($\ashift=0.024$) and run~B with $\beta_0=0.15$ at $\se=6$ ($\ashift=0.002$).

\begin{figure}[t]
\centering
\includegraphics[width=\textwidth]{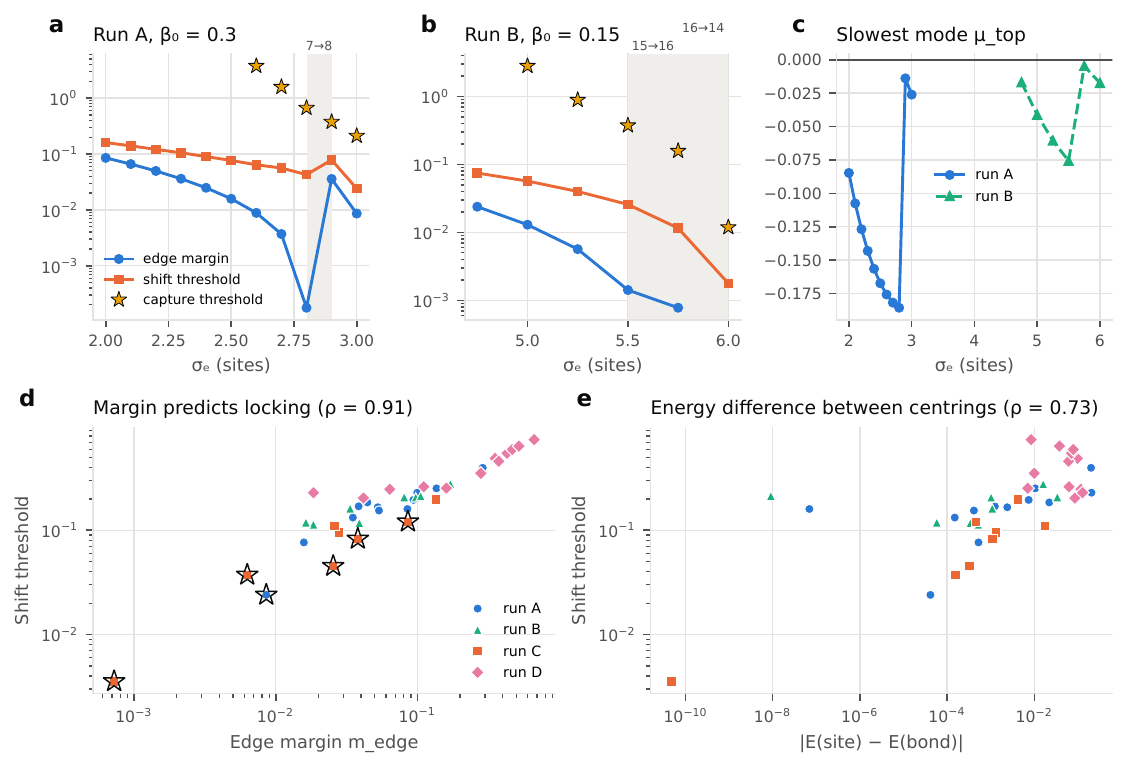}
\caption{\textbf{Locking is predicted by the edge unit's margin below threshold.} (a,b)~Fine sweeps across the loss of locking in run~A ($\beta_0=0.3$) and run~B ($\beta_0=0.15$), on a log scale. Three different quantities are plotted: the edge margin $m_{\mathrm{edge}}$, how far the unit just beyond the item's edge sits below threshold before any cue (blue circles); the shift threshold $\ashift$, the smallest edge input that displaces the item (orange squares); and the capture threshold $\acap$, the smallest cue at distance 2RF that drags the item to the cued location (stars, drawn only where capture occurs). Grey bands mark changes in item width. Within a fixed width the margin falls towards zero, locking weakens and capture appears; when the item recruits a unit, the margin resets. (c)~Slowest mode $\mutop$ in the same sweeps. It does not soften as locking is lost, and jumps when the width changes. (d)~Across all cells of runs A--D ($\se>0$; margin $>0$), the edge margin predicts the shift threshold (stars: cells with capture). (e)~The energy difference between the two centrings is a weaker predictor.}
\label{fig:locking}
\end{figure}

\paragraph{What predicts the barrier.} The energy framework of Section~\ref{sec:theory} offers a natural explanation of why a soft translation mode and a finite locking barrier can coexist. The energy is blind below threshold, so $\mutop$ describes the curvature of $\mathcal E$ for a fixed set of active units. Moving an item, however, requires changing that set: an inactive unit beyond the item's edge must be driven across threshold. That unit is held below threshold by its margin $m_{\mathrm{edge}}=-\sum_jW_{ej}y_j$ (Section~\ref{sec:recruit}). Across the 40 cells of runs A--D with $\se>0$ and a positive margin, $m_{\mathrm{edge}}$ predicted $\ashift$ with a rank correlation of 0.91 (Fig.~\ref{fig:locking}d). On the same cells, the energy difference between the item centred on a unit and centred between two units was a weaker predictor (0.73; Fig.~\ref{fig:locking}e). It was also sometimes vanishingly small where locking was strong: $7\times10^{-8}$ in run~A with $\beta_0=0.3$ at $\se=2$, where $\ashift=0.16$. Lattice locking is therefore not adequately characterised by the energy difference between alternative stationary representations, whereas the margin separating the edge unit from threshold strongly predicts the input needed to move an item and identifies the local mechanism of the threshold. Two caveats apply. Comparing the energies of two equilibria is not a calculation of the barrier along the transition path between them, on which other units also change; we do not claim that no such barrier exists. And a high rank correlation does not make the margin the only determinant: cue duration, connectivity and the rest of the active set also enter.

\paragraph{A fine sweep across the loss of locking.} To see what happens as locking is lost, we returned to the two regions in which it had vanished and varied $\se$ in finer steps: 0.1 in run~A ($\beta_0=0.3$, $\se=2$--3) and 0.25 in run~B ($\beta_0=0.15$, $\se=4.5$--6) (Fig.~\ref{fig:locking}a--c). The result has a sawtooth structure tied to the item's active set. While the item kept the same width (seven units in run~A for $\se=2.0$--2.8; 15 units in run~B for $\se=4.75$--5.5), its edge margin fell steadily towards zero (from 0.085 to 0.0002 in run~A), and the shift threshold fell with it. Within that range the margin ordered the shift threshold perfectly (rank correlation 1.0 in both runs), whereas the energy difference between centrings did not (0.15 in run~A). When the item recruited an additional unit ($\se=2.9$ in run~A), the margin jumped back up and locking strengthened, before falling again. Across all 17 cells of the fine sweep the rank correlation between margin and shift threshold was 0.95.

Notably, the slowest mode behaved quite differently. Within the seven-unit range of run~A, $\mutop$ became \emph{more} negative ($-0.085$ to $-0.186$) while locking weakened, and it jumped to $-0.014$ only when the item widened. Loss of locking is therefore not a smooth bifurcation of the stored state. It accompanies the approach of the edge unit to threshold, which is a change in the composition of the active set. The two quantities describe different things: $\mutop$ and $Q$ concern infinitesimal motion within the current active set, whereas a finite displacement by one site requires threshold crossings at the item's edges, which the linearisation does not see. The slow mode was translation throughout ($Q\ge 0.997$), except in the last run~B cell ($\se=6$), where the edge unit sits at threshold ($m_{\mathrm{edge}}=-0.0005$) and the slowest mode is a change at the edge rather than a shift ($Q=0.055$).

\subsection{Updating: costlier, repulsive, and finally captured}\label{sec:updating}

We next asked what happens when a new item arrives while one is already stored.

\paragraph{Adding an item becomes costly.} In the OCOS limit a second item can be added at distance $2\RF$ by a cue barely larger than the inhibition reaching that site (check C4): 0.010 to 0.031 across runs. As excitation widens the required cue rises steadily (Fig.~\ref{fig:continuum}f), to 0.70 in run~A ($\beta_0=0.2$, 35 times the narrow value) and 2.34 in run~B ($\beta_0=0.1$, 230 times). The reason is simple: wide items extend their inhibition farther, so a new item must overcome more of it.

\begin{figure}[t]
\centering
\includegraphics[width=\textwidth]{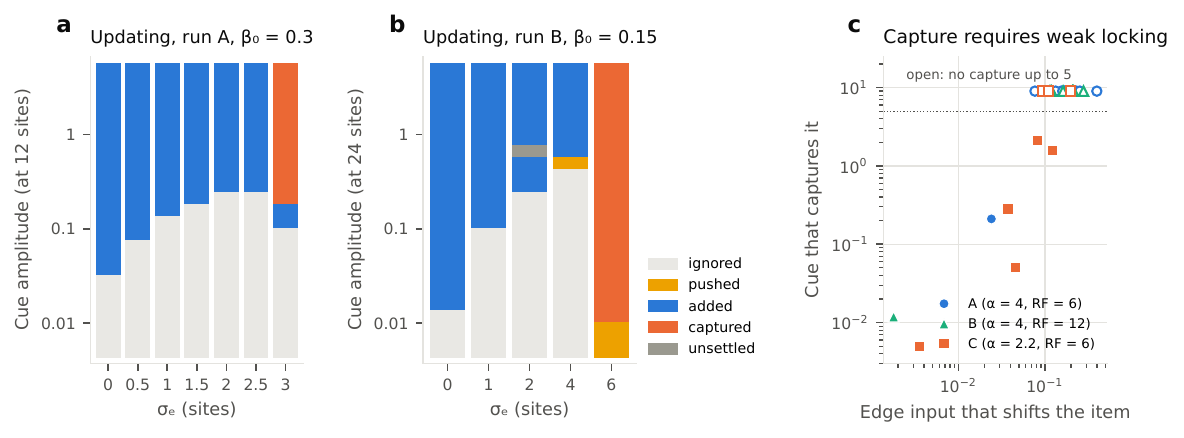}
\caption{\textbf{Updating changes character along the continuum.} (a,b)~Outcome of a cue at distance $2\RF$ from a stored item, by cue amplitude (log scale) and excitation width, for run~A ($\beta_0=0.3$) and run~B ($\beta_0=0.15$). Narrow and intermediate widths ignore weak cues and add an item for strong ones. At the widest setting the stored item is captured instead. (c)~Across all cells of runs A--C, the cue that captures an item against the edge input that shifts it (log--log). Open symbols at the top are cells in which no cue up to 5 caused capture.}
\label{fig:updating}
\end{figure}

\paragraph{Items repel.} When a second item is added, the two do not stay where they were cued (Fig.~\ref{fig:updating}). At the threshold cue, the stored item was pushed away from the new one by up to 7.5 sites in run~A and 10.3 sites in run~B, and the new item settled up to 11.7 and 28.0 sites beyond its cue. Subthreshold cues also displaced wide items without creating a new one (``pushed'' in Fig.~\ref{fig:updating}b). In the narrow limit neither effect occurs, since a stored single-unit item cannot move and a new item forms exactly at its cue.

\paragraph{Without locking, updating becomes capture.} In the two cells that lost their locking, updating changed character (Fig.~\ref{fig:updating}a,b). In run~A ($\beta_0=0.3$, $\se=3$), cues of 0.12--0.21 added a second item, but any cue of 0.21 or more dragged the stored item onto the cued location. In run~B ($\beta_0=0.15$, $\se=6$), even a cue of 0.005 pushed the item, and every cue of 0.012 or more captured it; no amplitude added a second item. In other words, the new input overwrites the memory instead of joining it.

Capture was not confined to these two cells. Across all 29 cells of runs A--C in which locking could be measured, capture occurred in seven, and all seven had $\ashift\le 0.121$ (Fig.~\ref{fig:updating}c). No cell with stronger locking showed capture at any amplitude up to 5. Among the capture cells, weaker locking went with a cheaper capture (rank correlation 0.82, $n=7$). Weak locking was, however, necessary but not sufficient: six cells with $\ashift$ between 0.076 and 0.119 showed no capture within the tested range. The fine sweeps of Section~\ref{sec:locking} show what is missing. Capture first appeared when the edge margin fell to about 0.01 ($\se=2.6$ in run~A, $m_{\mathrm{edge}}=0.009$; $\se=5.0$ in run~B, $m_{\mathrm{edge}}=0.013$), and its threshold then fell steeply as the margin approached zero (from 3.75 to 0.67 in run~A and from 2.81 to 0.012 in run~B; Fig.~\ref{fig:locking}a,b). Locking, in other words, is what lets a stored item hold its place against a new one, and it is strongly predicted by how close the item's edge is to recruiting another unit.

\subsection{Noise: drift is the width-dependent failure}\label{sec:noise}

\begin{figure}[t]
\centering
\includegraphics[width=\textwidth]{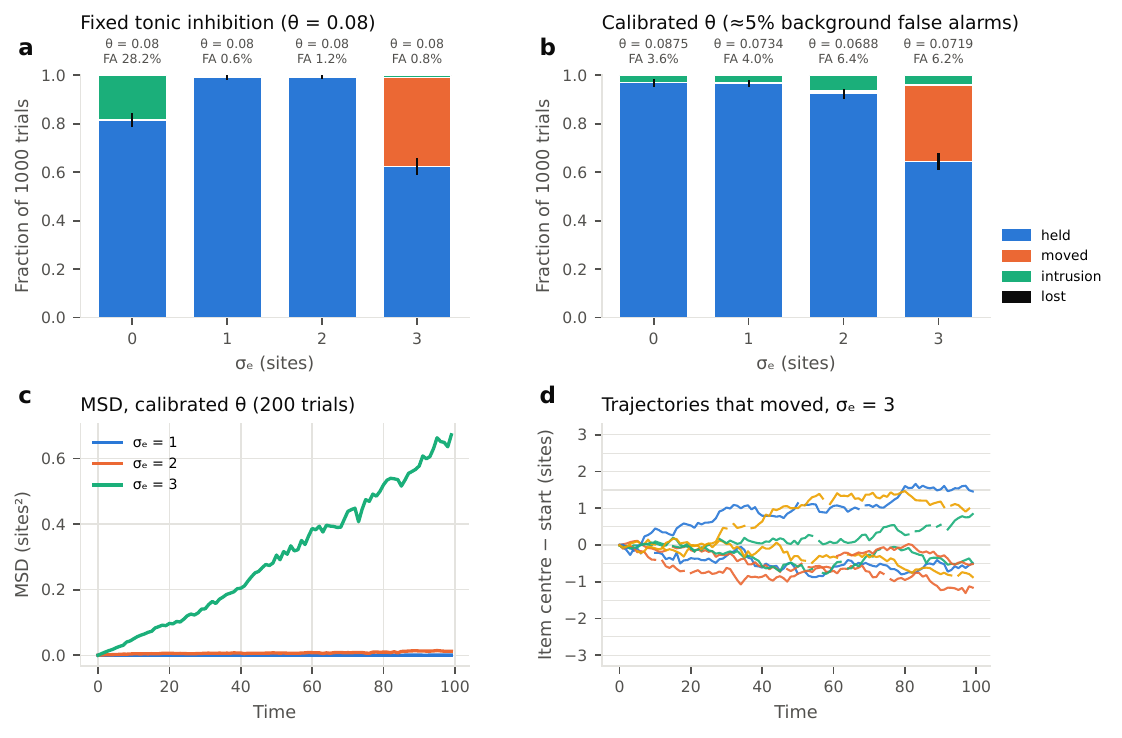}
\caption{\textbf{Under approximately matched background false alarms, drift is the width-dependent failure} (run~A configuration, $\beta_0=0.2$, $\sigma=0.03$, 1000 trials per condition). (a)~Fixed tonic inhibition $\theta=0.08$; above each bar, $\theta$ and the false-alarm rate of a network without a stored item. (b)~$\theta$ calibrated per width to give 5\% background false alarms. Error bars: 95\% Wilson intervals for the fraction held. (c)~Mean squared displacement of the item's centre, from 200 stored trajectories per width. (d)~Example trajectories at $\se=3$ that moved by at least 0.75 sites; grey lines mark site and bond positions.}
\label{fig:noise}
\end{figure}

Finally, we asked how memories fail under noise. Comparisons across widths face a confound, however. Units far from a stored item need some protection against recruitment (Proposition~\ref{prop:recruit}c), and how much a given tonic inhibition $\theta$ provides depends on width. Without a stored item, $\theta=0.08$ let a spurious item appear in 28\% of trials at $\se=0$ but in at most 1.2\% at $\se\ge 1$, because a smaller self-weight makes isolated units harder to recruit. We therefore compared single items (run~A configuration, $\beta_0=0.2$, noise $\sigma=0.03$, 1000 trials per condition) under two conditions: the same $\theta=0.08$ at every width, and a $\theta$ calibrated per width so that a network without a stored item acquired a spurious item in about 5\% of trials, so that background false alarms were approximately matched (achieved rates 3.6--6.4\%; calibrated $\theta$ between 0.069 and 0.088). Figure~\ref{fig:noise} shows the results with 95\% confidence intervals.

\paragraph{Intrusions track the background, not width.} With fixed $\theta$, the narrow network acquired spurious items in 18.4\% of trials (95\% CI 16.1--20.9\%), against at most 1\% at larger widths. With calibrated $\theta$, intrusion rates were 3.0--6.4\% at every width, close to the background false-alarm rate itself. The excess of intrusions at the narrow end under fixed $\theta$ therefore reflects how well the background is quieted, not a property of the stored item.

\paragraph{Drift grows with width.} Under both conditions the stored item never moved at $\se\le 1$ (0 of 1000 trials), moved in 1.0\% of trials at $\se=2$ (calibrated), and moved in 36.7\% (fixed; CI 33.8--39.7\%) and 31.5\% (calibrated; CI 28.7--34.4\%) of trials at $\se=3$. The mean squared displacement of the item's centre grew roughly linearly with time at $\se=3$, giving a finite-time effective diffusion coefficient of 0.003 sites$^2$ per unit time (an estimate over the simulated interval, not evidence of an asymptotic diffusion process in this finite, locked system), and was negligible at smaller widths (Fig.~\ref{fig:noise}c). The movement was graded rather than a sequence of hops between lattice positions. At $\se=3$, about 56\% of sampled centre positions lay more than 0.1 site from the nearest site or bond position (Fig.~\ref{fig:noise}d), consistent with the soft translation mode of Section~\ref{sec:continuum}. Over the delays simulated, then, the failure that depends on width is drift, which is the characteristic weakness of a bump attractor.

\subsection{When does widening produce bump-like behaviour?}\label{sec:controls}

\begin{figure}[t]
\centering
\includegraphics[width=\textwidth]{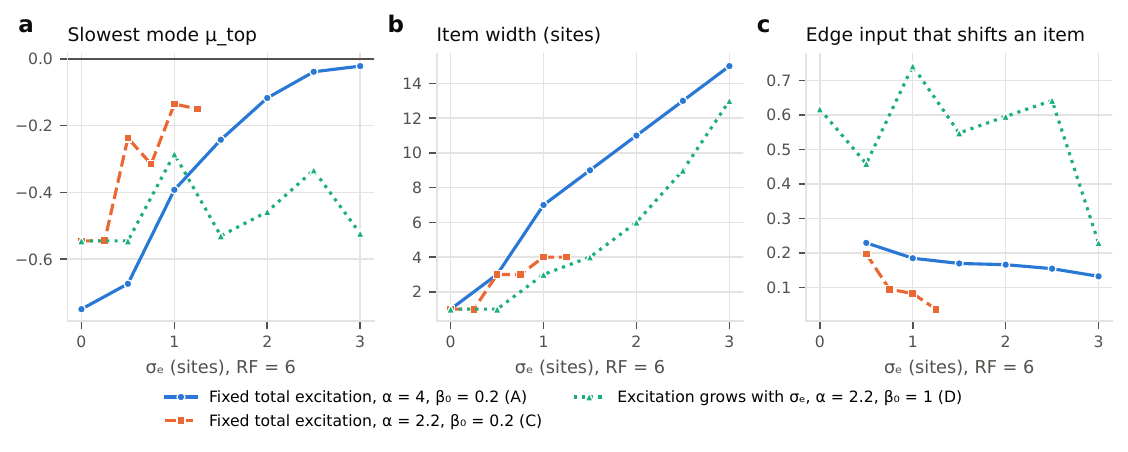}
\caption{\textbf{Bump-like softening appears only when total excitation is constrained and self-excitation is strong} ($\RF=6$). Run~A ($\alpha=4$, fixed total) against run~C ($\alpha=2.2$, fixed total) and run~D ($\alpha=2.2$, self-weight fixed so total excitation grows with $\se$). (a)~Slowest mode of one item. (b)~Item width. (c)~Edge input needed to shift an item.}
\label{fig:controls}
\end{figure}

One may ask whether bump-like behaviour follows from widening excitation as such. Two controls show that, in the regimes examined, it depends both on how excitation is widened and on the strength of self-excitation (Fig.~\ref{fig:controls}). The family itself is continuous in $\se$ under either normalisation; what differs is whether widening softens the translation mode.

\paragraph{When total excitation grows, items stay locked.} In run~D excitation was widened without normalisation, so that the self-weight stayed at $\alpha$ and total excitation grew with $\se$. Items still widened (to 13 units at $\se=3$, $\beta_0=1$), but they stayed deeply stable ($\mutop$ between $-0.28$ and $-0.55$) and strongly locked ($\ashift$ between 0.23 and 0.74). No cell showed capture, and items were held in at least 98\% of noisy runs at every width ($\sigma=0.05$, $\theta=0.15$). Adding a second item still became costlier, from 0.074 to 1.56, and at $\se=3$ no cue up to 5 succeeded. Thus, allowing excitatory mass to grow widens items while preserving strong attraction and lattice locking.

\paragraph{When self-excitation is weak, items stay narrow.} In run~C, with mass normalisation but $\alpha=2.2$ \citep{EndressSzabo2020}, items stayed narrow (at most four units by $\se=1.25$), and their slowest mode did not approach zero smoothly. Locking weakened quickly, however (from 0.20 at $\se=0.5$ to 0.037 at $\se=1.25$ with $\beta_0=0.2$, and to 0.004 with $\beta_0=0.3$), and five of the run's cells showed capture. At this level of excitation, widening produces the updating weakness of a bump attractor without its continuity. With noise, items did not survive tonic inhibition of 0.05 beyond $\se=0.5$.

\section{Discussion}\label{sec:discussion}

\subsection{One dial, two kinds of memory, and their price}

In the present study, we took the OCOS network that we had previously used to model enumeration and capacity limits \citep{Sengupta2014,Sengupta2017,Sengupta2025} and asked what happens to it as recurrent excitation is allowed to spread, with total excitation held fixed. OCOS and bump-attractor networks are usually treated as different models. Our results show that they can be placed in one family, in which a single parameter, the width of recurrent excitation, moves a network from spatially discrete towards distributed representations. That excitation width shapes bump width, stability and interference is not new in itself; neural-field models have long studied such dependencies \citep{Amari1977,Compte2000}. What is new here is the combination: an exact theory of the narrow end, a controlled path from it towards distributed excitation, and a comparison along that path of local stability, resistance to displacement and the outcome of updating.

The approach parallels our earlier account of enumeration, in which the subitizing and estimation ranges were not assumed a priori but emerged from one network as the strength of inhibition varied \citep{Sengupta2014,Sengupta2017}. Here, in the same spirit, the contrast between discrete and graded memory is not built into two architectures but emerges from one as excitation widens. What changes along the way can be summarised as a set of trade-offs between stability and cost.
\begin{enumerate}[leftmargin=*]
\item \emph{Continuity against cost.} A continuous, bump-like item occupies many units, sustains several times the activity of a single-unit item, and recovers from a perturbation tens of times more slowly (Section~\ref{sec:cost}). In this family, continuity is the one thing that widening buys, and every circuit cost we measured is higher than at the narrow end.
\item \emph{Capacity against interference range.} Narrow networks hold many items because each item occupies one unit and interacts with others only through inhibition (Proposition~\ref{prop:pair}). Wide networks hold few, because each item spreads and the separation two items need grows with how far they spread (Fig.~\ref{fig:continuum}e).
\item \emph{Stability against flexibility.} The slowest mode of a stored item, its translation, approaches zero as excitation widens (Fig.~\ref{fig:continuum}a), and at the widest settings noise moves the item (Fig.~\ref{fig:noise}). In narrow networks the stored item is never moved by noise.
\item \emph{Ease of adding against ease of overwriting.} Adding a second item is cheap in narrow networks and costly in wide ones, where items also repel. Where locking is lost, a new input no longer joins the memory but replaces it (Fig.~\ref{fig:updating}).
\item \emph{Locking against capture.} Items stay locked to the lattice along most of the family, because moving them requires recruiting an edge unit. Where the edge unit's margin vanishes, the same mechanism that held the item in place lets a new input drag it away (Fig.~\ref{fig:locking}).
\end{enumerate}
A recurring question when working-memory models are compared is which of their properties are essential and which incidental. Here the answer is specific: the properties that distinguish the two families follow from one structural parameter, and each can be traced to a mechanism, namely inhibition-limited interactions, spreading excitation, or threshold-induced locking.

The question of cost deserves a comment, because it is easy to confuse with the cost of simulation. Operation counts per integration step do not separate the regimes: the recurrent sum is a convolution and can be computed in $O(N\log N)$ operations at any width. The costs that do separate them are those the circuit itself pays, in units, activity, recovery time, spacing and the input needed to update. Framed this way, stability and cost are not independent properties of a model but two readings of the same parameter. A circuit could in principle choose its position along the family according to what a task requires: a narrow, cheap and robust store when many discrete items must be kept apart, and a wider, costlier one when what matters is a graded, smoothly shifting estimate of a single feature.

\subsection{Energy, thresholds and locking}

The energy formalism of \citet{Sengupta2024} does two jobs in this work. At the narrow end, its node-wise and pairwise criteria become the diagonal and the exact conditions of a single Hessian test (Corollary~\ref{cor:levels}). This accounts for the gap between them and, together with Proposition~\ref{prop:asym}, yields a capacity that is exact over all stable states. Along the family, the same energy provides a dynamical interpretation of why lattice locking and a soft translation mode coexist. Because the energy is blind below threshold, it can be nearly flat along translation (small $|\mutop|$, tiny energy differences between centrings) while moving the item still requires driving an inactive edge unit across threshold. Locking is therefore better understood as a recruitment threshold than through energy differences between item positions, and it is strongly predicted by the edge unit's margin (Section~\ref{sec:locking}). Capture appears where that margin vanishes. Whether a barrier exists along the full transition path between two item positions is a separate question that we have not addressed. We expect the same separation, between the curvature of an energy within an active set and the cost of changing the active set, to hold in other threshold networks.

\subsection{Discrete and continuous attractors}

The broader idea that discreteness trades against drift is not new. Heterogeneous excitation in a ring network stabilises a finite set of bump positions, slowing diffusion at the cost of coarse-graining what can be stored \citep{Kilpatrick2013,KilpatrickErmentrout2013}; homeostatic scaling \citep{Renart2003} and short-term facilitation \citep{Itskov2011} have been proposed as other ways to slow drift; and discrete and continuous attractors have been contrasted as models of graded persistent activity \citep{Brody2003}. Our results extend that picture in three directions. First, they concern multiple items, so capacity, interference range and updating enter alongside drift. Second, in a threshold network the lattice itself provides the discreteness, through the active set of each item. This is why lattice locking survives even as the translation mode softens: the slowest mode and the shift threshold measure different things, namely the curvature of the energy within an active set and the margin that must be overcome to change it (Section~\ref{sec:locking}). Third, the OCOS end of the family has an exact theory (Section~\ref{sec:theory}), so the continuum is anchored at one end by closed forms rather than only by simulation.

\subsection{Implications for visual working memory}

A hallmark of a useful computational model is that it explains known phenomena and also offers testable predictions. We therefore consider what the family implies for visual working memory.

\paragraph{Slot-like and resource-like regimes.} The OCOS end of the family has slot-like dynamics: items are discrete, all-or-none, and limited in number by a closed-form capacity (Theorem~\ref{thm:sym}). The bump end has resource-like dynamics: items are graded, interact over a range, and drift. These are dynamical regimes, not implementations of the behavioural theories. Slot and resource models make specific predictions about recall precision, guessing and set-size effects \citep{ZhangLuck2008,BaysHusain2008,Ma2014}, and testing those would require a behavioural read-out that we have not simulated. With that caveat, the contrast need not be a choice between architectures; it can be a position along a single circuit parameter. Spiking-network work has made a related point, showing a transition from distributed resources to limited slots under normalisation \citep{Wei2012}. The present family adds a specific, analysable mechanism for where along that axis a circuit sits.

\paragraph{Capacity depends on spacing, not only on number.} Under distance-dependent inhibition, how many items fit depends on how far apart they are (Proposition~\ref{prop:pair}; Fig.~\ref{fig:continuum}e). This leads to a direct prediction: in tasks where items occupy a common feature or spatial map, capacity estimates should rise with the spacing between items, and the spacing at which they stop interfering should scale with the width of the underlying representation. Topographic attractor models have already linked memory precision to the distances between remembered locations \citep{Almeida2015}.

\paragraph{Repulsion and overwriting.} In the wide regime a newly added item pushes a stored one away (Fig.~\ref{fig:updating}), which predicts repulsive biases between remembered items at separations just beyond the interference range. Where locking is weak, a strong new input overwrites rather than adds. In a working-memory task this would appear as a newly presented item displacing the memory of an earlier one at a nearby location, so that the earlier item is reported at the new location. Whether real circuits sit in this regime is an empirical question, but the model specifies the signature that would reveal it: a capture threshold that falls as resistance to displacement falls.

\paragraph{Total loss at high load.} Catastrophic interference in global-inhibition networks \citep{EndressSzabo2020} is, with this activation function, a long transient: silence is an unstable equilibrium rather than an attracting one (Proposition~\ref{prop:dip}). Whether it is observed in a trial therefore depends on how unequal the items' inputs are and on how long the delay is, not only on how many items there are.

\paragraph{Alternative accounts.} One alternative would be that discrete and graded memories are carried by distinct networks, for instance a capacity-limited priority map, of the kind OCOS networks were introduced to model \citep{Sengupta2014,Sengupta2017,Knops2014}, alongside feature-tuned bump attractors of the kind proposed for prefrontal cortex \citep{Compte2000,Wimmer2014}. Such an account is entirely possible, and our results do not rule it out. It is, however, less parsimonious: it must specify separately why the discrete store resists updating and why the graded store drifts, whereas in the present family both follow from one parameter and from a single locking mechanism. A second alternative is that resistance to displacement simply reflects the energy difference between item positions, as in a pinned continuous attractor. Across the parameter settings we examined, that difference was a weaker predictor than the edge margin, and it was sometimes negligible where locking was strong (Section~\ref{sec:locking}).

\subsection{Limitations}

The model is a one-dimensional ring of rate units. Visual working memory involves multiple feature dimensions and binding \citep{Ma2014}, and two-dimensional maps may change how items pack. We varied excitation width at two total-excitation levels and with one activation function; the threshold in $F$ is what produces lattice locking, and a smooth activation would be expected to reduce it. The background-quieting mechanism (tonic inhibition) was chosen for simplicity; intrusion rates depend on it, and for that reason we compared widths at approximately matched background false-alarm rates. The noise study covers one parameter path and delays of 100 time units. Noise was additive and white, whereas noise in real circuits has structure and intrinsic limits \citep{Burak2012}. The fine sweeps of Section~\ref{sec:locking} follow two paths through parameter space; a full map of where locking is lost, as a function of $\se$, $\beta_0$ and $\RF$, remains to be done. All simulations use a finite lattice, and we make no claim about convergence to the continuum neural-field limit. Finally, the closed forms of Section~\ref{sec:theory} hold exactly only for the OCOS limit. Along the rest of the family the results are numerical; Theorem~\ref{thm:energy} still holds there, but the closed-form capacities and separations do not.

\subsection{Future directions}

The most promising aspect of this line of work, in our view, is that it begins to connect questions that are usually studied separately: stability, capacity, interference, updating and drift. Several extensions follow naturally. In the present work, excitation width and self-excitation were fixed within a trial. Allowing self-excitation to vary in time, for example rising during stimulus presentation and decaying afterwards, would let the same network pass through competitive and maintenance regimes within a trial, and would link encoding and maintenance within a single recurrent architecture. Similarly, in our enumeration work we anticipated a mechanism for adjusting inhibition from trial to trial \citep{Sengupta2017}; an analogous adjustment of excitation width would allow a circuit to move along the family according to task demands. Finally, the predictions above, concerning spacing-dependent capacity, repulsion and capture, can be tested behaviourally, and the transition from locking to capture offers a candidate signature for neural recordings during memory maintenance. Understanding how a circuit balances stability against flexibility in this way would help bridge the gap between abstract computational models and the neural mechanisms of visual working memory.

\subsection{Conclusion}

In the present study, we showed that a single structural parameter, the width of recurrent excitation, carries an OCOS network towards a weakly stabilised, bump-like regime when total excitation is held fixed, and that it sets the price of continuity. Along the way the network gives up capacity, a short interference range and resistance to overwriting, and pays for continuity in units, activity and recovery time, while becoming vulnerable to drift. The narrow end has an exact theory of capacity, interference and robustness, built on the energy function of \citet{Sengupta2024}. Along the family, the translation mode softens while items stay locked, because moving an item requires recruiting a unit at its edge, a threshold that the energy within the active set does not register. When that threshold vanishes, updating turns into capture. Overall, these are properties of a circuit rather than of a modelling tradition, and they make predictions that can be tested.

\section*{Declarations}
\addcontentsline{toc}{section}{Declarations}

\paragraph{Funding.} No external funding was received for this work.

\paragraph{Competing interests.} The authors declare no competing interests.

\paragraph{Ethics approval and consent.} Not applicable; no human or animal data were collected.

\paragraph{Data and code availability.} The simulation code, the scripts that compute the theory checks and build every figure, and the resulting data files are provided as ancillary files with this arXiv submission (folder \texttt{anc/}; see \texttt{anc/README.md}).

\appendix
\section{Proofs}\label{app:proofs}

Throughout, $y_i=F(x_i)$, $F'(x)=(1+x)^{-2}$ for $x>0$ and $F'(x)=0$ for $x<0$. For $x>0$ we have $F(x)<x$ and the inverse $x=y/(1-y)$.

\subsection*{Theorem~\ref{thm:energy}}
\emph{(a)} Since $\Phi'(x)=x/(1+x)^2=xF'(x)$ for $x>0$ (and both sides vanish for $x<0$), and $W$ is symmetric,
\[
\frac{\partial\mathcal E}{\partial x_i}=F'(x_i)\Big[\lambda x_i-I_i-\sum_jW_{ij}y_j\Big]=-F'(x_i)\,\dot x_i ,
\]
so $\dot{\mathcal E}=\sum_i(\partial\mathcal E/\partial x_i)\dot x_i=-\sum_iF'(x_i)\dot x_i^2\le 0$. Equation~\eqref{eq:energy} is the Cohen--Grossberg function \citep{CohenGrossberg1983} for $a_i\equiv 1$, $b_i(x)=-\lambda x+I_i$, $c_{ij}=-W_{ij}$ and $d_j=F$.

\emph{(b)} Inactive units contribute nothing to $\mathcal E$ (since $\Phi=F=0$ there), and on the active set, in the coordinates $y_i$,
\[
\mathcal E=\sum_{i\in\A}\Big[\lambda\big(-\ln(1-y_i)-y_i\big)-I_iy_i\Big]-\tfrac12\sum_{i,j\in\A}W_{ij}y_iy_j ,
\]
because $\ln(1+x)=-\ln(1-y)$ and $1/(1+x)=1-y$. Then $\partial\mathcal E/\partial y_i=\lambda x_i-I_i-\sum_jW_{ij}y_j=-\dot x_i$ and $\partial^2\mathcal E/\partial y_i\partial y_j=\lambda(1-y_i)^{-2}\delta_{ij}-W_{ij}=H_{ij}$. For the Jacobian, $\partial\dot x_i/\partial x_j=-\lambda\delta_{ij}+W_{ij}F'(x_j)$. Columns belonging to inactive units have $F'=0$, so, ordering units as (active, inactive), the Jacobian is block lower-triangular with an inactive block $-\lambda I$. Its spectrum is that of the active block together with $-\lambda$. The active block is $J_\A=-\lambda I+W_{\A\A}D=-HD$, where $D=\diag(F'(x_i))$ and we used $\lambda(1+x_i)^2F'(x_i)=\lambda$. Since $D\succ 0$, $J_\A$ is similar to $-D^{1/2}HD^{1/2}$, which is symmetric and congruent to $-H$. Hence all eigenvalues of $J_\A$ are real, and by Sylvester's law of inertia \citep{HornJohnson2012} $J_\A$ has as many negative, zero and positive eigenvalues as $H$ has positive, zero and negative ones. If $H\succ 0$, every eigenvalue of the Jacobian is negative, and the principle of linearised stability gives local exponential stability. If $H$ has a negative eigenvalue, the Jacobian has a positive one and the equilibrium is unstable. If $H$ is singular, the Jacobian has a zero eigenvalue and higher-order terms decide. \qed

\subsection*{Proposition~\ref{prop:alpha}}
For an active unit at equilibrium, $\lambda x_i=\alpha y_i-\sum_{j\ne i}\beta_{ij}y_j\le\alpha y_i<\alpha x_i$, so $\lambda<\alpha$. Conversely, if $\alpha>\lambda$ then $x_w=\alpha/\lambda-1>0$ solves $\lambda x=\alpha x/(1+x)$, and every other unit rests at $x_l=-\beta_{lw}y_w/\lambda<0$ (for $\beta_{lw}>0$). Its Hessian is the scalar $\lambda(1+x_w)^2-\alpha=\alpha(\alpha-\lambda)/\lambda>0$. \qed

\subsection*{Theorem~\ref{thm:sym} and Corollary~\ref{cor:wta}}
With $S$ equally active units, an active unit satisfies $\lambda x=\alpha y-\beta(S-1)y=A_Sy$, so $1+x_S=A_S/\lambda$. A positive solution exists if and only if $A_S>\lambda$. An inactive unit satisfies $\lambda x_l=-\beta Sy_S<0$. On $\A$,
\[
H=\Big(\tfrac{A_S^2}{\lambda}-\alpha-\beta\Big)I+\beta\,\mathbf 1\mathbf 1^{\top},
\]
whose eigenvalues are $A_S^2/\lambda-(\alpha+\beta)$ on the $(S-1)$-dimensional subspace $\mathbf 1^{\top}v=0$, and $A_S^2/\lambda-\alpha+(S-1)\beta=A_S(A_S-\lambda)/\lambda$ along $\mathbf 1$. The second is positive exactly when the state exists; for $S\ge 2$, the first is positive exactly when $A_S^2>\lambda(\alpha+\beta)$. With $D=(\lambda/A_S)^2I$ and $W_{\A\A}=(\alpha+\beta)I-\beta\mathbf 1\mathbf 1^{\top}$, $J_\A=-\lambda I+W_{\A\A}D$ gives \eqref{eq:modes}. Because $A_S>0$, the stability condition is $A_S>\sqrt{\lambda(\alpha+\beta)}$, which rearranges to \eqref{eq:Sst}. $\Sst<\Smax$ because $\alpha+\beta>\lambda$ implies $\sqrt{\lambda(\alpha+\beta)}>\lambda$. Solving $(\alpha-(S-1)\beta)^2=\lambda(\alpha+\beta)$ as a quadratic in $\alpha$ gives roots $\tfrac12[2(S-1)\beta+\lambda\pm\sqrt{\lambda^2+4\lambda S\beta}]$. At the smaller root $A_S<0$, so the relevant condition is $\alpha>\alpha_S$, the larger root \eqref{eq:alphaS}. For $S=1$ there is no transverse subspace and stability reduces to existence, $\alpha>\lambda$.

For the corollary, $\alpha_S(\beta)$ increases with $S$, so $\alpha\le\alpha_2$ implies $\alpha\le\alpha_S$ for every $S\ge 2$. All multi-item equal-activation states are then unstable, while the single item is stable for every $\alpha>\lambda$. \qed

\subsection*{Proposition~\ref{prop:asym}}
Write $u_i=1+x_i$ and $c=\alpha+\beta$. With $y=1-1/u$, an active unit at equilibrium satisfies $\lambda x_i-c\,y_i=-\beta Y$, with $Y=\sum_{j\in\A}y_j$, that is,
\[
\lambda u_i+\frac{c}{u_i}=K,\qquad K=\lambda+c-\beta Y,
\]
so every active $u_i$ is a root of the same quadratic $\lambda u^2-Ku+c=0$. Active units therefore take at most two values, $u_l\le u_h$, and if they differ then $u_lu_h=c/\lambda$. Write $d_i=\lambda u_i^2-c$, so that $H=\diag(d_i)+\beta\mathbf 1\mathbf 1^{\top}$. $H$ is a positive rank-one update of $\diag(d_i)$, so by eigenvalue interlacing \citep{HornJohnson2012} the second-smallest $d_i$ is at least the smallest eigenvalue of $H$. If $H\succ 0$, at most one $d_i$ is non-positive.

Suppose a stable equilibrium had unequal activations. Using $u_lu_h=c/\lambda$ and $\Delta=u_h-u_l>0$, we get $d_l=\lambda u_l(u_l-u_h)=-\lambda u_l\Delta<0$ and $d_h=\lambda u_h\Delta>0$. Since at most one $d_i$ is non-positive, exactly one unit takes the value $u_l$ and $S-1$ units take $u_h$. With exactly one negative diagonal entry, $H$ has at most one non-positive eigenvalue, so $H\succ 0$ if and only if $\det H>0$. Since $\det H=\prod_i d_i\,(1+\beta\sum_i d_i^{-1})$ and $\prod_i d_i<0$, this requires
\[
\frac{\beta}{\lambda u_l\Delta}>1+\frac{\beta(S-1)}{\lambda u_h\Delta}.
\]
Multiplying by $\lambda u_lu_h\Delta>0$ and using $\lambda u_lu_h=c$ gives $\beta u_h>c\Delta+\beta(S-1)u_l$. Substituting $u_h=u_l+\Delta$, this becomes $-\alpha\Delta>\beta(S-2)u_l$. The left side is negative and the right side is non-negative for $S\ge 2$, a contradiction. Hence every stable equilibrium has equal activations. \qed

\subsection*{Proposition~\ref{prop:dip} (sketch)}
For identical units and inputs the equations are invariant under permutations, so the subspace of equal activations is invariant. On it, $\dot x=-\lambda x+A_SF(x)<0$ for $x>0$ when $A_S\le\lambda$, so $x\to 0^+$; for $S>\Smax$ ($A_S<\lambda$) this decay is eventually exponential. Near $x=0^+$, $F(x)\approx x$, and the linearised dynamics of the active units are $\dot x_i=(\alpha-\lambda)x_i-\beta\sum_{j\ne i}x_j$. The uniform mode has rate $\alpha-\lambda-(S-1)\beta=A_S-\lambda<0$, and every transverse mode has rate $\alpha-\lambda+\beta>0$. A relative asymmetry $\varepsilon$ therefore grows relative to the mean at rate $(\alpha+\beta-\lambda)+(\lambda-A_S)=S\beta$ and becomes of order one after $\ln(1/\varepsilon)/(S\beta)$. Over that time the mean has decayed by a factor $\varepsilon^{(\lambda-A_S)/(S\beta)}$, which gives \eqref{eq:dip}. Once differences are of order one, the leading units suppress the others and recover. The offset $t_0$ is the time to approach the linear regime from the initial state. \qed

\subsection*{Proposition~\ref{prop:pair}}
With two active units and inhibition $b=\beta_{ij}$ between them, an active unit satisfies $\lambda x=(\alpha-b)y$. On the two active units, $H$ has diagonal $(\alpha-b)^2/\lambda-\alpha$ and off-diagonal $b$, with eigenvalues $(\alpha-b)^2/\lambda-\alpha\pm b$. The smaller is positive if and only if $(\alpha-b)^2>\lambda(\alpha+b)$, that is, $b^2-(2\alpha+\lambda)b+\alpha^2-\lambda\alpha>0$. The larger root of this quadratic exceeds $\alpha$ (giving $\alpha-b<0$), so the condition is $b<\beta^*$ as in \eqref{eq:betastar}. Existence ($\alpha-b>\lambda$) follows, since $(\alpha-b)^2>\lambda(\alpha+b)>\lambda^2$. Other units receive inhibition from both items and are inactive. With $\beta_{ij}=\beta_0e^{-d^2/2\RF^2}$, the condition $\beta_{ij}<\beta^*$ is $d>d^*$, and the smallest integer $d$ satisfying it is $\lfloor d^*\rfloor+1$ (at least 1). \qed

\subsection*{Proposition~\ref{prop:recruit}}
With fixed inhibition $m>0$ and $I_l=0$, unit $l$ obeys $\dot x=-\lambda x-m$ for $x\le 0$, so it rests at $-m/\lambda$, and $\dot x=h(x)-m$ for $x>0$. Here $h(0)=0$, $h'(0)=\alpha-\lambda>0$ and $h''<0$; $h$ is maximised where $(1+x)^2=\alpha/\lambda$, with maximum $(\sqrt\alpha-\sqrt\lambda)^2$. If $m\ge h_{\max}$, $\dot x\le 0$ for all $x>0$ and the unit returns below threshold. Otherwise $h(x)=m$ has two positive roots $x_u<x_s$, $\dot x>0$ on $(x_u,x_s)$, and a kick that carries the unit from $-m/\lambda$ above $x_u$ sends it to $x_s$. If $m=0$, then $x_u=0$. \qed

\section{The shunting OCOS network}\label{app:shunting}

In the shunting variant of the OCOS network \citep{Grossberg1973,CohenGrossberg1983}, excitation and inhibition are gated by the unit's own activation:
\begin{equation}\label{eq:shunt}
\dot x_i=-x_i+(B-x_i)F(x_i)-(D+x_i)\sum_{k\ne i}F(x_k)+I_i ,\qquad B,D>0 .
\end{equation}
It also has Cohen--Grossberg form, with $a_i(x)=D+x>0$ on $(-D,\infty)$, so Theorem~\ref{thm:energy}(b) carries over: at a nondegenerate equilibrium $J_\A=-\diag(a_i)\,H\,\Lambda$ with $\Lambda=\diag(F'(x_i))$ and $H$ the energy Hessian, which is similar to $-S H S$ with $S=\diag(a_iF'(x_i))^{1/2}$. The energy criterion and linear stability therefore again coincide, in the regular region where $a_i>0$, $F'(x_i)>0$ on the active set and no unit sits at threshold.

\begin{proposition}\label{prop:shunt}
Assume $I=0$, an equilibrium in the regular region ($x_i>-D$ for all units, active units strictly above threshold and inactive units strictly below it), and stability in the strict linear sense of Theorem~\ref{thm:energy}. Then a state with $n$ equally active units has $x[n]=\big(B-1-(n-1)D\big)/(n+1)$. It exists if and only if $c_n=B-(n-1)D>1$, that is, $n<n_{\max}=1+(B-1)/D$. Inactive units decouple, and the active modes are
\[
\mu_\parallel=-\frac{(n+1)x[n]}{1+x[n]}<0,\qquad
\mu_\perp=-1-\frac{n\,x[n]}{1+x[n]}+\frac{B+D}{(1+x[n])^2}.
\]
The state is therefore stable if and only if $c_n(c_n+n)>(B+D)(n+1)$.
\end{proposition}

\begin{proof}
The equilibrium follows from multiplying the fixed-point equation by $1+x$. Differentiating \eqref{eq:shunt} at the symmetric state gives a diagonal entry $-1-y+(B-x)y'-(n-1)y$ and an off-diagonal entry $-(D+x)y'$ among active units, with $y'=(1+x)^{-2}$. The identities $1+x[n]=(c_n+n)/(n+1)$, $1+(n+1)x[n]=c_n$ and $B-(n-1)D-nx[n]=1+x[n]$ give the two modes and the stability condition.
\end{proof}

Gating bounds activity to $(-D,B)$, but it does not remove the existence limit $n_{\max}$, so shunting does not by itself protect against loss of items at high load. For $(B,D)=(3,0.2)$, $(4,0.5)$, $(2.5,0.1)$ and $(6,0.3)$, the largest stable $n$ is 3, 3, 4 and 6, against $n_{\max}=11$, 7, 16 and 17.7. In each case the condition above agreed with the numerically computed Jacobian at every $n<n_{\max}$.

\section{Numerical methods}\label{app:methods}

\paragraph{Integration.} Equation~\eqref{eq:model} was integrated with the Euler method, step $0.01$, and with the Euler--Maruyama method for noise, adding $\sigma\sqrt{0.01}\,\xi_i$ per step with independent standard normal $\xi_i$. Items were created by a cue of amplitude 1 at one unit for 5 time units, followed by 60 time units without input.

\paragraph{Stability gate and symmetry breaking.} A state was accepted as a memory state only if its largest residual $|\dot x_i|$ was below $10^{-3}$ and every eigenvalue of its Jacobian was negative. A state that failed was kicked by $10^{-4}$ along its leading eigenvector, plus small random components, on active units only, and allowed to settle for a further 60 time units; this was repeated up to six times. Restricting the kick to active units matters: by Proposition~\ref{prop:recruit}(c), a kick to a distant quiescent unit would create a spurious item. The gate is needed for two reasons. A single-unit cue is mirror-symmetric, so the dynamics can remain on a symmetric saddle indefinitely (for instance, an item centred on a unit whose stable form is centred between two units). And near the bump-attractor end the relaxation time $1/|\mutop|$ can exceed the settling time. Without the gate, interference ranges and capacities at large $\se$ were overestimated, in one case by a factor of three.

\paragraph{Counting items.} For $\se>0$ an item is a contiguous run of active units. When neighbour excitation is below 1\% of self-excitation ($\se<0.33$) every active unit is counted as its own item, because adjacent single-unit items would otherwise merge into one cluster. Item centres are $F$-weighted circular means.

\paragraph{Measures.} The interference range $\dmin$ is computed from simultaneous cues at separations $d=1,2,\dots$; it is the smallest $d$ from which no larger tested separation ends in merging or the loss of an item. Pairs that had not settled into a stable state were counted separately and not treated as failures. Updating cues were applied at a fixed distance $2\RF$ for 10 time units, at 25 log-spaced amplitudes between 0.005 and 5. Outcomes need not be monotone in amplitude, so the whole grid was classified, and the first ``added'' boundary was refined by local bisection. For the noise runs, ``moved'' means a displacement of at least 0.75 sites: an item centred between two units can be tipped half a site to one side, which is not a genuine move. The effective diffusion coefficient is half the slope of the mean squared displacement of the item's centre over time, computed from runs that kept exactly one item.

\paragraph{Slow mode, energy and edge margin.} The translation overlap is $Q=|v\cdot t|/(\|v\|\,\|t\|)$, where $v$ is the right eigenvector of the active Jacobian belonging to $\mutop$ and $t_i=(x_{i+1}-x_{i-1})/2$ is the discrete derivative of the item profile, both restricted to active units. The energy \eqref{eq:energy} was evaluated at two mirror-symmetric states: an item cued at one unit (centred on a unit) and an item cued equally at two adjacent units (centred between them). Without kicks the dynamics stay in the corresponding symmetric subspace, so each state is an equilibrium of the full network, whether or not it is stable; both were settled for 200 time units. The edge margin is $m_{\mathrm{edge}}=-\lambda x_e$ at the relaxed item, where $e$ is the first unit beyond the item's edge (the unit at which the shift probe is applied).

\paragraph{Circuit costs.} For each cell of runs A and B a single item was formed as in Section~\ref{sec:measures}. We recorded its number of active units, its total output $\sum_iF(x_i)$ and its slowest relaxation time $\tau=1/|\mutop|$ (\texttt{cost\_analysis.py}). The remaining costs ($\dmin$, capacity and $\aadd$) come from the main sweep.

\paragraph{Fine transition sweep.} Run~A with $\beta_0=0.3$ and $\se=2.0,2.1,\dots,3.0$, and run~B with $\beta_0=0.15$ and $\se=4.5,4.75,\dots,6.0$. Each cell was analysed with the same measures as the main sweeps, except interference range and capacity.

\paragraph{Noise study.} Run~A configuration with $\beta_0=0.2$, $\se\in\{0,1,2,3\}$, noise $\sigma=0.03$, and 1000 trials of 100 time units per condition. In the \emph{fixed} condition $\theta=0.08$ in every cell. In the \emph{calibrated} condition $\theta$ was chosen per cell, by bisection over $[0,0.2]$ with 200 trials per evaluation, so that a network without a stored item acquired a spurious item in 5\% of trials. The achieved false-alarm rate was then re-estimated from 500 independent trials. Proportions are reported with Wilson 95\% confidence intervals. Item-centre trajectories were sampled once per time unit.

\paragraph{Checks.} Before every sweep the code runs the five checks of Table~\ref{tab:checks} at $\se=0$ and refuses to proceed if any fails. A separate script reproduces every closed form of Section~\ref{sec:theory} against simulation (Fig.~\ref{fig:theory}).

\begin{table}[h]
\centering\small
\caption{Parameters of the runs. All runs: $\lambda=1$, Euler step 0.01, cue amplitude 1 for 5 time units, settling 60 time units, updating cue at $2\RF$ for 10 time units, noise runs of 100 time units (100 per cell in runs A--D, 1000 per condition in the noise study).}
\label{tab:params}
\begin{tabular}{@{}l l c c c l l l@{}}
\toprule
Run & Excitation & $\alpha$ & $N$ & $\RF$ & $\beta_0$ & $\se$ & Noise ($\sigma$; $\theta$) \\
\midrule
A & fixed total & 4 & 160 & 6 & 0.2, 0.3 & 0, 0.5, \dots, 3 & 0.03; 0.08, 0.1 \\
B & fixed total & 4 & 320 & 12 & 0.1, 0.15 & 0, 1, 2, 3, 4, 6 & 0.03; 0.05, 0.08 \\
C & fixed total & 2.2 & 160 & 6 & 0.2, 0.3 & 0, 0.25, \dots, 1.25 & 0.02; 0.05, 0.08 \\
D & fixed self-weight & 2.2 & 160 & 6 & 0.3, 0.6, 1.0 & 0, 0.5, \dots, 3 & 0.05; 0.15 \\
\midrule
Fine A & fixed total & 4 & 160 & 6 & 0.3 & 2.0, 2.1, \dots, 3.0 & -- \\
Fine B & fixed total & 4 & 320 & 12 & 0.15 & 4.5, 4.75, \dots, 6.0 & -- \\
Noise & fixed total & 4 & 160 & 6 & 0.2 & 0, 1, 2, 3 & 0.03; 0.08 / calibrated \\
\bottomrule
\end{tabular}
\end{table}

\paragraph{Software.} Python 3 with NumPy for simulation and Matplotlib for figures. The sweep in run~A takes about 2 minutes and run~B about 6 minutes on a single CPU core without noise. Noise runs add several minutes per cell.

\end{document}